\documentclass{article}
\usepackage{caption}
\usepackage{indentfirst}
\usepackage{subcaption}
\usepackage{amsmath}
\usepackage{amssymb}
\usepackage{authblk}
\usepackage[english]{babel}
\usepackage{blindtext}
\usepackage{authblk}
\usepackage{setspace}
\usepackage{graphicx}
\usepackage{titling}
\usepackage{blindtext}
\usepackage{natbib}
\usepackage{float}
\usepackage{algorithm,algorithmic}
\usepackage[colorlinks=true,linkcolor=blue,urlcolor=black,bookmarksopen=true,citecolor=blue]{hyperref}
\usepackage{setspace}
\AtBeginDocument{%
  \addtolength\abovedisplayskip{-0.18\baselineskip}%
  \addtolength\belowdisplayskip{-0.18\baselineskip}%
}

\title{Inference for Log-Gaussian Cox Point Processes using Bayesian Deep Learning: Application to Human Oral Microbiome Image Data}
\author{Shuwan Wang$^1$, Christopher K. Wikle$^2$, Athanasios C. Micheas$^2$,\\ Jessica L. Mark Welch$^3$ Jacqueline R. Starr$^{4,5}$, Kyu Ha Lee$^1$}
\date{%
    $^1$Harvard T.H. Chan School of Public Health, Boston, MA, U.S.A.\\%
    $^2$Department of Statistics, University of Missouri-Columbia, Columbia, MO, U.S.A.\\%
    $^3$The Forsyth Institute, Cambridge, MA, U.S.A.\\%
    $^4$Channing Division of Network Medicine, Brigham and Women's Hospital, Boston, MA, U.S.A.\\%
    $^5$Department of Medicine, Harvard Medical School, Boston, MA, U.S.A.
}

\begin{document}

\maketitle

\begin{abstract}
It is common in nature to see aggregation of objects in space. Exploring the mechanism associated with the locations of such clustered observations can be essential to understanding the phenomenon, such as the source of spatial heterogeneity, or comparison to other event generating processes in the same domain. Log-Gaussian Cox processes (LGCPs) represent an important class of models for quantifying aggregation in a spatial point pattern. However, implementing likelihood-based Bayesian inference for such models presents many computational challenges, particularly in high dimensions.  In this paper, we propose a novel likelihood-free inference approach for LGCPs using the recently developed \textit{BayesFlow} approach, where invertible neural networks are employed to approximate the posterior distribution of the parameters of interest. \textit{BayesFlow} is a neural simulation-based method based on "amortized" posterior estimation. That is, after an initial training procedure, fast feed-forward operations allow rapid posterior inference for any data within the same model family. Comprehensive numerical studies validate the reliability of the framework and show that \textit{BayesFlow} achieves substantial computational gain in repeated application, especially for two-dimensional LGCPs. We demonstrate the utility and robustness of the method by applying it to two distinct oral microbial biofilm images.   

\end{abstract}


\section{\textbf{Introduction}}
In nature, species commonly form groups where conditions are best for survival, e.g., plants might cluster together where soil nutrients are rich. Different species may cluster similarly or in different ways on a spatial domain. In some cases a species may enable another's survival, for example by providing nutrients, physical scaffolding, or protection. Similarly, in communities of organisms, spatial organization may reflect or drive the key components that maintain balance in the micro-environment. Characterizing and quantifying the spatial distributions of species is crucial for understanding the underlying biological mechanisms in many disciplines. However, this can be a difficult problem since spatial point pattern data are often incomplete and/or aggregated. Even after taking into account the impacts of candidate covariates, there will often be a residual component of spatial correlation, requiring the introduction of latent spatial stochastic processes in the model. \par

The log Gaussian Cox process (LGCP), introduced in \cite{moller1998log}, provides a flexible and intuitive framework for quantifying spatial distributions associated different event generating processes (such as bacterial taxa) and for addressing a wide range of scientific questions in spatial and spatio-temporal point pattern data. As an extension of the non-homogeneous Poisson process, the LGCP considers an intensity function that is both location specific and also stochastic, since it incorporates a latent Gaussian process (GP) that determines the spatial structure of clusters. 
\par
Although LGCPs offer an appealing modeling approach, performing likelihood-based inference is computationally challenging due to the intractable nature of the likelihood function, which involves a stochastic integral. Many approaches have been proposed to overcome these computational obstacles. A widely used strategy is discretization, where  a set of representative points approximates the integral using a discrete Riemann sum (e.g., \citealp{moller1998log}; \citealp{moller2003statistical}). Yet, this discretization leads to another computational complexity, as likelihood evaluation requires inverting large variance-covariance matrices associated with the latent GP. High-dimensional LGCP inference is particularly demanding, and overcoming these computational burdens has been an active area of research. Existing solutions, such as  nearest neighbor GPs, Vecchia approximations, and reduced-rank representations, provide alternatives but often trade computational efficiency for accuracy \citep[e.g., see][for an overview and comparison]{heaton2019case}.

Inference for LGCPs can be performed from a frequentist or Bayesian perspective \citep[e.g., see][for recent overviews]{teng2017bayesian, dovers2023fast}. Bayesian inference, in particular, is computationally demanding, requiring sampling-based methods such as Markov chain Monte Carlo (MCMC) or Hamiltonian Monte Carlo (HMC) approaches. The Metropolis-adjusted Langevin algorithm (MALA) approach is a classic Bayesian approach to fit LGCPs \citep{moller1998log}, while the integrated nested Laplace approximation (INLA) \citep{rue2009approximate} has also proven effective, leveraging Gaussian Markov random field to reduce computational complexity. Despite these advances, Bayesian inference for LGCPs remains computationally expensive, leading to a growing interest in approximate inference methods. For example, likelihood-free approaches, such as approximate Bayesian computation (ABC), offer a flexible simulation-based strategies for carrying out approximate inference in a Bayesian framework \citep[e.g., see][]{beaumont2010approximate}. This method circumvents the need for evaluating a likelihood function by comparing  observed and simulated data using summary statistics \citep{vihrs2022approximate}. However, constructing suitable summary statistics is non-trivial, especially for models with intractable likelihoods. Other simulation-free methods have been proposed as an alternative to ABC. These methods estimate the likelihood and perform inference by comparing various estimating functions \citep{moller2007modern}. For instance, in case of minimum contrast estimation (e.g., \citealp{moller1998log}), the optimal parameter estimate minimizes the distance between a non-parametric estimate of a second-order summary statistic and its theoretical expression. Similarly, \cite{guan2006composite} proposes a composite likelihood approach based on an estimating function relying on the second-order intensity function. However, in these two-stage approaches, parametric bootstrap simulations used to quantify uncertainty of parameter estimates often result in an underestimation of their variance \citep{waagepetersen2016analysis}.

\par
Deep learning techniques can transform estimation problems into optimization problems, for which we have computationally efficient algorithms (e.g., stochastic gradient descent). Recently, neural network-based methods have been extensively applied to model spatial/spatio-temporal data sets (see the review in \citealp{wikle2023statistical}). In particular, these approaches have been used with generative models that have intractable likelihoods, such as the neural-based estimation of max-stable processes for modeling spatial extreme events \citep{lenzi2023neural}. Other implementations consider the use of convolutional neural networks to learn the likelihood function of a spatial process through a specifically constructed binary classifier \citep{walchessen2024neural}, the construction of ``neural Bayes" estimators to approximate Bayes estimators \citep{sainsbury2024likelihood}, and further extension of the latter for irregular spatial data by using graph neural networks \citep{sainsbury2024neural}. While these methods are solely for parameter point estimation, approaches have been developed to quantify the associated uncertainty (\citealp{lenzi2023neural}, \citealp{sainsbury2024likelihood} and \citealp{walchessen2024neural}). 
Importantly, these approaches for neural estimation are all examples of so-called ``amortized inference.'' That is, one invests a significant upfront computational expense to train flexible deep neural models on a class of models for which simulated data can easily be generated across a wide variety of model parameters. Then, given new data, estimation and inference from this trained model is obtained very fast. Hence, the cost of the initial training phase is ``amortized over time”  \citep[see][for a recent overview]{zammit2024neural}. 

\par
Our proposed methodology builds on a related line of research in which variational inference is achieved through the construction of chains of invertible transformations via deep neural networks, allowing for the representation of arbitrarily complex posterior distributions (e.g., \citealp{rezende2015variational}; \citealp{kingma2016improved}; \citealp{radev2020bayesflow}). As in the aforementioned neural estimation approaches, \textit{BayesFlow} \citep{radev2020bayesflow} is also based on amortized estimation. In an initial training phase, the neural network learns a complex mapping to be used for inference. \textit{BayesFlow} then performs a rapid inference phase through efficient feed-forward operations. Once the neural network is trained, not only can posterior parameter inference efficiency be improved tenfold or even hundredfold than classical Bayesian sampling methods, but the trained network can also be reused for inferences with new data from the same model family. 

\par
In this paper, we develop a novel likelihood-free inference method for LGCPs using \textit{BayesFlow}. Invertible neural networks (INNs) are adopted for the approximation of the posterior distribution of the parameters of interest. There are several key benefits to this approach. First, the proposed framework is flexible and general, enabling Bayesian inference without the need to evaluate the likelihood function. Second, instead of deriving point-wise estimation of parameters such as proposed in \cite{vihrs2022using}, our method relies entirely on the INN for statistical inference. This allows for the direct derivation of posterior distributions rather than a single point-estimator, facilitating the straightforward quantification of uncertainty. Finally, the framework does not require fixing any test data sets for evaluation during the procedure. Once the neural network is trained, our method can infer full posteriors on any data set within the same model family. 

\par
This paper proceeds as follows. In Section 2, we introduce the \textit{BayesFlow} framework and specific considerations for using this algorithm to model LGCPs. In Section 3, we present the results of simulation studies that illustrate the effectiveness of the proposed method. In Section 4, we apply our approach to microbial biofilm image data, illustrating its practical utility. Finally, in Section 5, we provide concluding remarks and discuss possible future extensions.

\section{\textbf{Methodology}}

In this section, we introduce the likelihood-free method \textit{BayesFlow} to perform simulation-based inference for LGCPs. The background of LGCPs is provided in Section 2.1 and the essential ideas behind \textit{BayesFlow} are described in Section 2.2. The structure of requisite INNs is introduced in Section 2.3. In Section 2.4, we discuss the process of validating the selection of priors and then provide the amortized inference algorithm for LGCPs with \textit{BayesFlow} in Section 2.5.   

\textbf{2.1 LGCP Model}

Following \cite{moller1998log}, the random intensity function $\lambda$ of a LGCP at a spatial location $\boldsymbol{s}$ in a bounded region $\textit{W} \subset R^d$ is defined as
\begin{align}
\lambda(\boldsymbol{s})=exp(Z(\boldsymbol{s})), \hspace{0.2in} \boldsymbol{s} \in \textit{W}, \tag{1}
\end{align}
where $\{ Z(\boldsymbol{s}) \}_{\boldsymbol{s} \in \textit{W}}$ is a GP with constant overall mean $\mu \in \textit{R}^d$ and exponential covariance function 
\begin{align}
c(\boldsymbol{s_i},\boldsymbol{s_j})=\sigma^2 exp(-||\boldsymbol{s_i}-\boldsymbol{s_j}||/\rho), \hspace{0.2in} \boldsymbol{s_i}, \boldsymbol{s_j}\in \textit{W}. \tag{2}
\end{align}
Here, $\sigma^2$ is the variance and $\rho > 0$ is a spatial range parameter of the latent GP. Each realization of a LGCP on the observation window $\textit{W}$ with intensity specified in Eq.(1) depends on the parameter vector of interest $\boldsymbol{\theta}= (\mu, \rho, \sigma^2)'$. Accordingly, the likelihood function of the observed point pattern data $\boldsymbol S=\{{\bf s}_1,\dots,{\bf s}_N\}$ can be written as 
\begin{align}
L(\boldsymbol{S}|\lambda(\boldsymbol{s})) \propto exp(\hspace{0.02in} -\int_\textit{W}\lambda(\boldsymbol{s})d\boldsymbol{s})\prod_{i=1}^N\lambda(\boldsymbol{s_i}). \tag{3}
\end{align}
The presence of the stochastic integral in Eq.(3) brings computational challenges in performing Bayesian inference for parameter vector $\boldsymbol{\theta}$ since the resulting likelihood is analytically intractable \citep{moller1998log}.

\textbf{2.2 Likelihood-free Approach: \textit{BayesFlow}}


\textit{BayesFlow} \citep{radev2020bayesflow} is a fully likelihood-free approach that can be used to directly infer the posterior distribution of the underlying parameters for a model of interest $\boldsymbol{\theta}$ given observations (i.e., an observed point patten) $\boldsymbol S$, i.e., $p(\boldsymbol\theta|\boldsymbol S)$. We focus on illustrating the essential ideas behind the framework here, and refer the reader to \cite{radev2020bayesflow} for more details (e.g., proof of the learned posterior distribution under perfect convergence of the proposed neural network).
In short, rather than requiring the evaluation of the intractable likelihood function in Eq.(3), this approach requires only the ability to generate samples (i.e., point patterns) from the model, which can be performed through simulation given a set of data-generating parameters $\boldsymbol{\theta}$, i.e., $\boldsymbol S^{(i)} \sim p(\boldsymbol S|\boldsymbol{\theta})$.

The ultimate goal of \textit{BayesFlow} is to train a conditional invertible neural network (cINN), $p_{\boldsymbol{\phi}}(\boldsymbol{\theta}|\boldsymbol S)$, to approximate the target posterior distribution as precisely as possible, i.e., $p_{\boldsymbol{\phi}}(\boldsymbol{\theta}|\boldsymbol S)  \approx p(\boldsymbol{\theta}|\boldsymbol S)$. Specifically, the approximate posterior $p_{\boldsymbol{\phi}}$ is constructed in terms of a cINN $f_{\boldsymbol{\phi}}$ with a normalizing flow \citep{rezende2015variational} implemented between $\boldsymbol{\theta}$ and a standard Gaussian latent variable $\boldsymbol{Y}$:
\begin{align*}
\boldsymbol{\theta}&= f^{-1}_{\boldsymbol{\phi}}(\boldsymbol{Y}; \boldsymbol S) \\
\boldsymbol{Y}&= f_{\boldsymbol{\phi}}(\boldsymbol{\theta}; \boldsymbol S),
\tag{4}
\end{align*}
where $f_{\boldsymbol{\phi}}: \mathbb{R}^D \rightarrow \mathbb{R}^D$ is an invertible function parameterized by a vector of neural-network parameters $\boldsymbol{\phi}$ 
(see the details about the formulation of $f_{\boldsymbol{\phi}}$ in Section 2.3) and $\boldsymbol{Y} \sim \mathcal{N}_D (\boldsymbol{0}, \mathbb{I}_D)$. 

To ensure that the output of $f^{-1}_{\boldsymbol{\phi}}(\boldsymbol{Y};  \boldsymbol S)$ represents the true posterior, $p(\boldsymbol\theta|\boldsymbol S)$, the loss function is based on  
minimizing the Kullback-Leibler (KL) divergence between the true and the approximate posterior for all possible point patterns $\boldsymbol S$, and is approximated by adopting the Monte Carlo estimate with a batch of $\textit{J}$ simulated data sets and data-generating parameters $\{ (\boldsymbol S^{(j)}, \boldsymbol{\theta}^{(j)}) \}_{j=1}^{\textit{J}}$ and by applying the change of variable rule of probability (see the derivation details in \citealp{radev2020bayesflow}). Specifically,  we minimize the objective
\begin{equation}
\begin{split}
\hat{\boldsymbol{\phi}} 
& = \underset{\boldsymbol{\phi}}{argmin} \hspace{0.04in} \frac{1}{\textit{J}} \sum_{j=1}^{\textit{J}} \hspace{0.03in} \left( \frac{||f_{\boldsymbol{\phi}}(\boldsymbol{\theta}^{(j)}; \boldsymbol S^{(j)})||^2}{2} \hspace{0.03in} - \hspace{0.03in} log \left| det J_{f_{\boldsymbol{\phi}}}^{(j)} \right|  \right),
\end{split}
\tag{5}
\end{equation}
where $J_{f_{\boldsymbol{\phi}}}^{(j)}$ represents $\partial f_{\boldsymbol{\phi}}(\boldsymbol{\theta}^{(j)}; \boldsymbol S^{(j)})/ \partial \boldsymbol{\theta}^{(j)}$ (the Jacobian of $f_{\boldsymbol{\phi}}$ evaluated at $\boldsymbol{\theta}^{(j)}$ and $\boldsymbol S^{(j)}$).  
The loss function in Eq.(5) not only minimizes the Kullback-Leibler (KL) divergence between the true and  approximate posterior, but also ensures that $\boldsymbol{Y}$ follows the prescribed standard Gaussian distribution. \par
There are several methods proposed in the literature to accommodate replicated data in neural Bayes estimation. For example, \citet{gerber2021fast} suggest averaging all the neural network estimators of each single realization or training a new network for a fixed number of realizations.  Alternatively, \citet{sainsbury2024likelihood} adopt a permutation-invariant neural network, and \citet{walchessen2024neural} evaluate the likelihood of arbitrary realizations through the product of a classifier. 
Importantly, \textit{BayesFlow} also accommodates the situation where arbitrary replicated data is available by suggesting an additional flexible summary network $h_{\boldsymbol{\psi}}(\boldsymbol S_{1:n}^{(j)})$ that can be substituted for $\boldsymbol S^{(j)}$ in Eq.(5) to incorporate such realizations within the objective when necessary \citep[see][for more details]{radev2020bayesflow}. 

\textbf{2.3 Structure of Invertible Networks}

\cite{radev2020bayesflow} use the affine coupling block (ACB) introduced by \cite{dinh2016density} when applying the cINN. An ACB implements an invertible nonlinear transformation. 
Specifically, each ACB involves four separate fully connected neural networks denoted as $s_1(.)$, $s_2(.)$, $t_1(.)$, and $t_2(.)$. Letting $\boldsymbol{x}$ be the input vector of $f_{\boldsymbol{\phi}}$ and $\boldsymbol{g}$ the output vector, the forward and inverse transformations yield $f_{\boldsymbol{\phi}}(\boldsymbol{x})=\boldsymbol{g}$ and $f_{\boldsymbol{\phi}}^{-1}(\boldsymbol{g})=\boldsymbol{x}$, respectively. The input vector $\boldsymbol{x}$ is divided into two halves (i.e., $\boldsymbol{x}= (\boldsymbol{x}_1, \boldsymbol{x}_2)$ with $\boldsymbol{x}_1=\boldsymbol{x}_{1:D/2}$ and $\boldsymbol{x}_2=\boldsymbol{x}_{D/2+1:D}$, where $D/2$ is a floor division) and the operations are implemented on each half of the input. \par
Here, we adopt an augmented version of ACB for implementation. That is, the internal neural networks for each ACB are augmented to use the summary statistics $\widetilde{\boldsymbol S}$ of the data as a conditional input to incorporate the summarized simulated or observed data. Specifically, the forward direction of each ACB gives
\begin{align}
\begin{cases}
\boldsymbol{g}_1 =\boldsymbol{x}_1 \odot exp(s_1(\boldsymbol{x}_2, \widetilde{\boldsymbol S}))+t_1(\boldsymbol{x}_2, \widetilde{\boldsymbol S}) \\
\boldsymbol{g}_2 =\boldsymbol{x}_2 \odot exp(s_2(\boldsymbol{g}_1, \widetilde{\boldsymbol S}))+t_2(\boldsymbol{g}_1, \widetilde{\boldsymbol S}),
\tag{6}
\end{cases}
\end{align}
and the inverse operation is obtained straightforwardly as
\begin{align}
\begin{cases}
\boldsymbol{x}_2 =(\boldsymbol{g}_2-t_2(\boldsymbol{g}_1, \widetilde{\boldsymbol S})) \odot exp(-s_2(\boldsymbol{g}_1, \widetilde{\boldsymbol S}))\\
\boldsymbol{x}_1 =(\boldsymbol{g}_1-t_1(\boldsymbol{x}_2, \widetilde{\boldsymbol S})) \odot exp(-s_1(\boldsymbol{x}_2, \widetilde{\boldsymbol S})),
\tag{7}
\end{cases}
\end{align}
where $\odot$ represents element-wise multiplication. \par
Splitting the input has multiple benefits. First, it ensures that the Jacobian of the transformation is a strictly upper or a lower triangular matrix, and thus, its determinant is very inexpensive to compute for the loss function in Eq.(5). Second, these internal neural networks are flexible and are not required to be invertible because they are evaluated only in the forward transformation during both the forward and inverse operations of each ACB. Last, splitting the input allows multiple ACBs to be built upon each other to increase the expressiveness of the nonlinear transformation, while the whole chain of transformations remains invertible.   
In summary, the entire cINN can be expressed as $\boldsymbol{Y}=f_{\boldsymbol{\phi}}(\boldsymbol{\theta}; \widetilde{\boldsymbol S})$, with the inverse operation $\boldsymbol{\theta}=f^{-1}_{\boldsymbol{\phi}}(\boldsymbol{Y}; \widetilde{\boldsymbol S})$.

\textbf{2.4 Prior Consideration}

Since each realization of a LGCP on the observation window $\textit{W}$ with intensity specified in Eq.(1) only depends on the parameter vector $\boldsymbol{\theta}= (\mu, \rho, \sigma^2)'$, we assign the following priors to the LGCP model parameters for a bounded window:
\begin{align*}
\mu &\sim Unif(3, 6)  \notag \\
\rho &\sim Unif (0, 0.15\times|W|) \notag \\
\sigma^2 & \sim Unif(0, 2),
\tag{8}
\end{align*}
where $Unif(a,b)$ is the uniform distribution on the interval $(a,b)$. We follow the suggested prior distributions adopted in \cite{vihrs2022approximate} for $\mu$ and $\sigma^2$. Computational expense will increase or decrease with the number of points in a simulated point pattern. Fewer points yield a sparser point pattern, which can also make inference more difficult. Recall that $exp(\mu+\sigma^2/2)$ is the expected number of points in a point pattern $\boldsymbol S$. Limiting the range of the prior distributions of $\mu$ and $\sigma^2$ helps ensure that simulated point patterns will not yield unreasonably too many or too few points. For $\rho$, we choose to truncate the prior distribution to the interval of over 15$\%$ of the image window for practical considerations given the observation window. \par
A predetermined bounded parameter space is a common assumption in contemporary simulation-based inference since simulating parameters from an unbounded space is computationally inefficient. We have described prior ranges that can be predetermined by practical considerations relevant to LGCPs. For other models where the parameters of interest do not have such limitations in practice or for which guidance is lacking, we recommend checking the \textit{prior predictive distribution} (see Chapter 6 in \citealt{wikle_spatio-temporal_2019}) over the specified priors to confirm the validity.

\textbf{2.5 Amortized Inference Algorithm for LGCP}

For most Bayesian inference algorithms, the estimation process must be repeated in its entirety with observations from a new data set. However, \textit{BayesFlow} implements amortized inference by separating the whole framework into an initial computationally costly training phase and a cheaper and efficient inference phase. Algorithm 1 summarizes the essential steps in the original \textit{BayesFlow} \citep{radev2020bayesflow}, employing an \textit{online learning} approach where data are simulated on demand. However, if the simulation brings a high computational cost, a more time-efficient \textit{offline learning} approach is recommended prior to the training, in which a fixed number of simulations are conducted and stored. \par
Since the likelihood for LGCP 
is not expressible in closed form, it is not obvious which summary statistics $\widetilde{\boldsymbol S}$ to use to evaluate the generated random point pattern $\boldsymbol S$. However, the parameters $\mu$ and $\sigma^2$ are expected to be closely related to $exp(\mu+\sigma^2/2)$, the expected number of points in a point pattern $\boldsymbol S$ \citep{moller1998log}. The empirical estimate of the L-function may also be considered, as it is a summary statistic commonly used to assess the degree of clustering. \cite{vihrs2022approximate} propose another set of summary statistics that may help capture the clustering behavior and spatial heterogeneity. Assume that the observation window $\textit{W}$ is split into $q^2$ squares. Then, letting $n(\boldsymbol S \cap \textit{W}_{i,j})$ be the number of points in $\boldsymbol S$ falling in $\textit{W}_{i,j}$ ($i,j=1,\dots,q$), we can adopt descriptive statistics associated with $n(\boldsymbol S \cap \textit{W}_{i,j})$ and that can be calculated for an arbitrary set of $q$ values. In summary, for a generated point pattern $\boldsymbol S^{(j)}$, we consider the following set of summary statistics as $\widetilde{\boldsymbol S}^{(j)}$ when implementing Algorithm 1:
\begin{enumerate}
\item  $n_{log}:= log(n(\boldsymbol S^{(j)}))$.
\item  $\hat{L}(r)-r$ : normalized L-function empirical estimates evaluated at distance $r$,\\
where $r$ is chosen to be $m$ uniformly spaced values between 0 and $0.2 \times|W|$.
\item  $P_{max,q}:= \underset{i,j=1,\dots,q}{max}(\{n(\boldsymbol S^{(j)} \cap \textit{W}_{i,j})/n(\boldsymbol S^{(j)})   \})$, \\
$P_{min,q}:= \underset{i,j=1,\dots,q}{min}(\{n(\boldsymbol S^{(j)} \cap \textit{W}_{i,j})/n(\boldsymbol S^{(j)})   \})$, \\
$P_{log var,q}:= log\left(var\left(\{n(\boldsymbol S^{(j)} \cap \textit{W}_{i,j})/n(\boldsymbol S^{(j)})\}_{i,j=1}^q\right)\right)$.
\end{enumerate}
\begin{algorithm}[H]

    \begin{algorithmic}[1]

    \STATE \textbf{Training phase} (online learning with batch size \textit{J}):
    \REPEAT
    \FOR{j = 1, \dots, \textit{J}}

        \STATE Sample model parameters from prior specified in Eq.(8): $\boldsymbol{\theta}^{(j)} \sim p(\boldsymbol{\theta})$.

        \STATE Run the specified model in Eq.(1) to create a synthetic observation: $\boldsymbol S^{(j)} \sim p(\boldsymbol S |\boldsymbol{\theta}^{(j)})$.
        \STATE Summarize the dataset $\boldsymbol S^{(j)}$ with summary statistics $\widetilde{\boldsymbol S}^{(j)}$.
        \STATE Pass $(\boldsymbol{\theta}^{(j)}, \widetilde{\boldsymbol S}^{(j)})$ through the inference network in forward direction: $\boldsymbol{Y}^{(j)} = f_{\boldsymbol{\phi}}(\boldsymbol{\theta}^{(j)}; \widetilde{\boldsymbol S}^{(j)})$.
    \ENDFOR
    \STATE Compute loss according to Eq.(5) from the training batch $\{ (\boldsymbol{\theta}^{(j)}, \widetilde{\boldsymbol S}^{(j)}, \boldsymbol{Y}^{(j)}) \}_{j=1}^{\textit{J}}$.
    \STATE Update neural network parameters $\boldsymbol{\phi}$ via backpropagation.
    \UNTIL convergence to $\hat{\boldsymbol{\phi}}$
    \STATE
    \STATE \textbf{Inference phase} (given observed or test data $\boldsymbol S^o$):
    \STATE Summarize the observed dataset $\boldsymbol S^o$ with summary statistics $\widetilde{\boldsymbol S}^o$.
    \FOR{l=1, \dots, \textit{L}}
        \STATE Sample a latent variable instance: $\boldsymbol{Y}^{(l)} \sim \mathcal{N}_D (\boldsymbol{0}, \mathbb{I}_D)$.
        \STATE Pass $(\widetilde{\boldsymbol S}^o, \boldsymbol{Y}^{(l)})$ through the inference network in inverse direction: $\boldsymbol{\theta}^{(l)}=f^{-1}_{\hat{\boldsymbol{\phi}}}(\boldsymbol{Y}^{(l)}; \widetilde{\boldsymbol S}^o)$.
    \ENDFOR
    \STATE Return $\{ \boldsymbol{\theta}^{(l)} \}_{l=1}^{\textit{L}}$ as a sample from $p(\boldsymbol{\theta}|\boldsymbol S^o)$
    \end{algorithmic}
\caption{Amortized Bayesian Inference for LGCPs With the \textit{BayesFlow} Method \citep{radev2020bayesflow}}
\end{algorithm}

\section{\textbf{Simulation study}}

In this section, we describe simulation studies that illustrate the effectiveness of \textit{BayesFlow} for inference for LGCPs. The specific experimental details are provided in Section 3.1. Transformation of priors for the parameters of interest is introduced in Section 3.2. In Section 3.3 we validate the performance of using \textit{BayesFlow} by comparing it with true posterior estimates from Markov chain Monte Carlo (MCMC) for realizations of LGCPs in one-dimension $\subset [0,1]$ (1-D LGCPs) and in the planar bounded window $\subset [0,1]^2$ (2-D LGCPs), respectively. 

\textbf{3.1 Experimental Details in Applying \textit{BayesFlow} for LGCPs}
 
In this experiment, we performed a total of $15,000$ and $10,000$ update iterations in the training phases for 1-D and 2-D LGCPs, respectively, with each iteration of batch size $\textit{J}=16$. For parameter vector $\boldsymbol{\theta}= (\mu, \rho, \sigma^2)'$, we used a \textit{BayesFlow} with 12 ACBs. Since the L-function is not well-defined in the 1-D case, we constructed similar sequenced statistics summarizing the proportion of pairs of points lying within $m$ uniformly spaced values of $r$ between 0 and $0.2 \times |W|$. Thus, for each simulated point pattern $\boldsymbol S$ in the 1-D LGCPs, we used a summary vector of size 59 (i.e., $n_{log}$, $m=40$ and $q=2, 3, 4, 5, 10, 20$). For each simulated point pattern $\boldsymbol S$ in the 2-D LGCPs, we applied inhomogeneous L-function estimates directly and used a summary vector of size 55 (i.e., $n_{log}$, $m=40$ and $q=2, 3, 4, 5, 10$). All networks were implemented and trained in R using the $\textit{torch}$ library via backpropagation. The stochastic gradient descent algorithm was implemented with a starter learning rate of $10^{-6}$ and a decay rate of $0.95$ at every $1,000^{th}$ iteration. To minimize training costs (in terms of time), an offline learning approach was used, where data were generated according to Eq.(1) prior to the training. The generalized loss from Eq.(5) decreased with increasing update steps consisting of i.i.d. draws of pairs of $(\boldsymbol{\theta}, \boldsymbol S)$ (i.e., simulation parameters and its corresponding synthetic data) until convergence. Once the networks converged, we saved the trained network parameters $\hat{\boldsymbol{\phi}}$ and applied them to perform the amortized inference. 

\textbf{3.2 Prior Transformations}

As introduced in Section 2.4, we assigned prescribed-range uniform distributions for the parameter vector $\boldsymbol{\theta}= (\mu, \rho, \sigma^2)'$ in the LGCPs for both the simulation studies and the application. To make sure the posterior draws for each parameter from the inference phase also lay within the restricted range imposed by the priors, we conducted logit transformations for the parameters in $\boldsymbol{\theta}$ after converting them to $Unif(0,1)$, respectively. We passed the logit transformed-$\boldsymbol{\theta}$ through the training phase and then back-transformed the posterior draws from the inference phase by $invlogit$ as well as linear transformation to the original prescribed scale to ensure valid posterior distribution.

\textbf{3.3 Validation Performance}

In this subsection, we demonstrate that the \textit{BayesFlow} method accurately recovers the parameters of a LGCP model with intractable likelihood by learning summary statistics from raw data. We evaluated the performance of the likelihood-free \textit{BayesFlow} approach by comparing it with true posterior draws using MCMC. Sampling a higher dimensional GP within the LGCP is computationally challenging using MCMC. Therefore, we computed validation metrics for both approaches over 300 datasets for 1-D LGCPs (in Section 3.3.1) and demonstrated both performance only on selective datasets for 2-D LGCPs (in Section 3.3.2). \par
We used the same prior distributions as specified in Eq.(8) and considered two metrics, the NRSSE (normalized root sum squared error) and $R^2$ (the coefficient of determination), as suggested in \citealp{radev2020bayesflow}. The NRSSE is conducted over $\boldsymbol{J}$ different sets of true parameter vectors $\{ \boldsymbol{\theta}^{(j)}\}_{j=1}^J$ to assess accuracy of point estimates $\{ \hat{\boldsymbol{\theta}}^{(j)}\}_{j=1}^J$ in recovering true parameter values and is given by:
\begin{align}
NRSSE=\sqrt{\sum_{j=1}^J \frac{({\theta}^{(j)}-\hat{{\theta}}^{(j)})^2}{\theta_{max}-\theta_{min}}}. \tag{9}
\end{align}
The NRSSE can be used to compare the recovery across parameters with various numerical ranges since it is scale-free on account of the normalization factor $\theta_{max}-\theta_{min}$. In addition to NRSSE, we also computed $R^2$ to assess the proportion of variance of the true parameters that is captured by their estimates. This allows for accuracy of point estimates to be evaluated across multiple sets of parameters.

\textbf{3.3.1 1-D LGCPs}

Based on the simulations, the training loss decreased until convergence (Figure 1, left). Besides inspecting the loss plot, we also validated and tested if the training objective was reached. Since the training objective assigned a standard Gaussian distribution to the latent variable $\boldsymbol{Y}$ in the loss function Eq.(5), we should expect that the latent space will exhibit such a distribution under good convergence. Indeed, during the training phase, a complex mapping of the data was learned so that inference could be made in this transformed space, meaning the training objective was reached (Figure 1, right). 

\begin{figure}[H]
 \centering
 \includegraphics[width=0.453\linewidth]{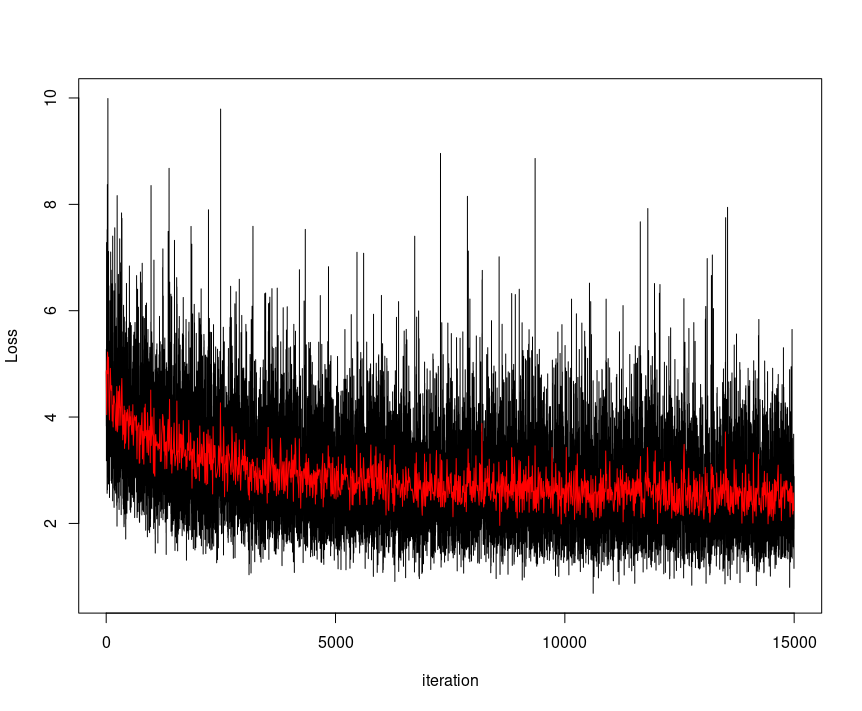}
 \includegraphics[width=0.453\linewidth]{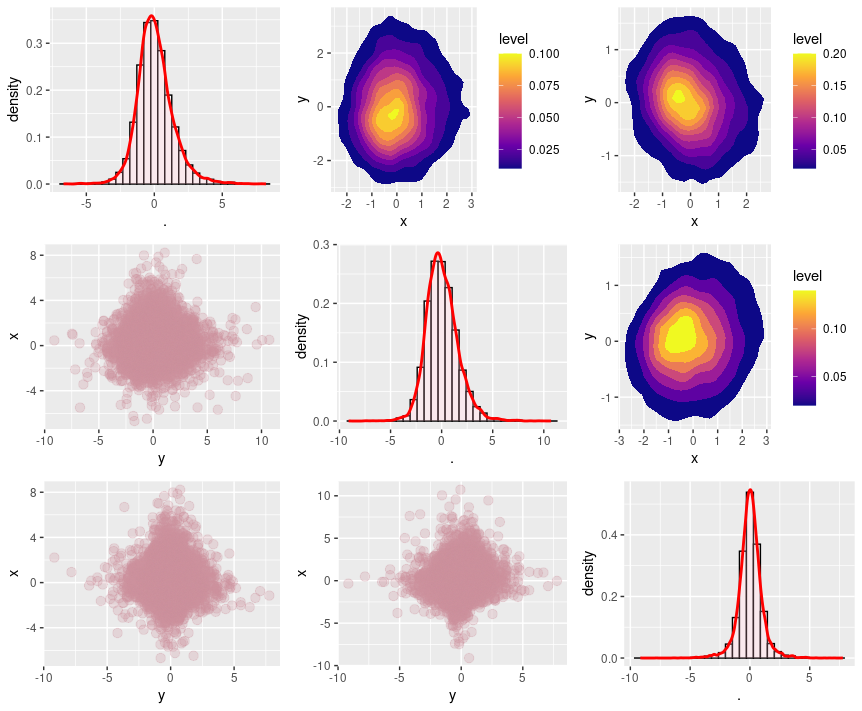}
 \caption{Inspection of the training loss and the latent transformed space $\boldsymbol{Y}$. Left: The training loss trace plot of 15,000 update iterations. The red line represents the moving average for every $10^{th}$ iteration; Right: Inspection of the transformed latent $\boldsymbol{Y}$-space.}
\end{figure}
To evaluate how inference performed, we implemented two \textit{BayesFlow} methods using either the original summary statistics or standardized summary statistics as conditional input. We compared the results of these methods with those of MCMC over 300 different sets of true parameters in the 1-D LGCPs. Compared with the original summary statistics, the standardized summary statistics yielded parameters much more similar to their true values, especially for $\mu$ and $\rho$ (Figure 2). These results demonstrate that the true parameters $\mu$ and $\sigma^2$ can be accurately recovered by the \textit{BayesFlow} method based on learning standardized summary statistics from raw data. 

The \textit{BayesFlow} method struggled to identify $\rho$ properly when the true parameter value was near 0 (the boundary of the parameter space). In this situation, the neural network approach tended to estimate $\rho$ to be near the mean of its prior distribution, i.e., the data were down-weighted compared to the prior since the spatial dependency was relatively weak in the resulting point pattern (the smaller $\rho$, the smaller covariance in Eq.(2)). 
In contrast, the MCMC performed better at detecting weaker dependency. This result agrees with \cite{zhang2005towards}, who pointed out that when the true range parameter $\rho$ was small relative to the sampling domain, it can be well estimated from the GP, yet when the true value of $\rho$ was greater than 0.10, the MCMC showed a significant tendency for underestimation, perhaps due to lack of identifiability \citep{zhang2004inconsistent}. However, the \textit{BayesFlow} method performed slightly better in this latter situation. For almost all test datasets, the posterior densities of $\sigma^2$ and $\rho$ spread over the entire prior range (high posterior variance), which agreed with the findings in \cite{zhang2004inconsistent}. They showed that for infill asymptotics (i.e., keeping the domain fixed and increasing the number of points within), both the range and variance parameters were inconsistent estimators for a Mat\'ern covariance Gaussian process. Inspecting the full posterior obtained by \textit{BayesFlow} for $\mu$, we observed that on most test datasets, its posterior 95$\%$ credible intervals were much wider compared with those from MCMC, indicating larger uncertainty in the obtained estimates. \cite{radev2020bayesflow} showed similar results for a different illustrative model in their simulations.

Based on NRSSE (the smaller, the better), the $\sigma^2$ parameter turned out to have the largest uncertainty, regardless of method of inference (Table 1). As pointed out in \cite{ying1991asymptotic}, for a 1-D GP with an exponential covariance function, neither the $\sigma^2$ nor $\rho$ parameters can be estimated consistently given that the process was observed in the unit interval. This might explain why the estimates had such large variation in the posterior densities (Figure 2). 

In terms of both the NRSSE and $R^2$ metrics, the \textit{BayesFlow} method using standardized summary statistics reasonably recovered the point estimates via posterior means  for $\mu$ and $\sigma^2$ (Table 1). It accomplished this nearly 10 times faster than MCMC, which was set to run for 30,000 iterations for each simulated dataset of the 1-D LGCP to ensure posterior convergence and proper posterior inference (Table 1). Even though the upfront training cost was nearly 12.5 hours, the extra effort of learning a global \textit{BayesFlow} model upfront was worthwhile if one had multiple instances from the same model, since the network had to be trained only once.

\textbf{3.3.2 2-D LGCPs}

In the classical geostatistical context, it is well known that it can be difficult to estimate the correlation range parameter, $\rho$, and this difficulty is exacerbated in the high-dimensional spatial point process setting \citep{diggle2013spatial}. In addition, it can be quite expensive to estimate GPs in high-dimensional settings without special parameterizations \citep[e.g.,][]{banerjee2017high}. Thus, in comparing validation performance and metrics between \textit{BayesFlow} and MCMC on simulated 2-D LGCP datasets, we did so only on datasets in which the resulting point patterns had adequate information for MCMC to draw inference on the latent Gaussian process. MCMC was set to run 50,000 iterations for each simulated dataset to ensure posterior convergence and proper posterior inference. The upfront training cost of learning the \textit{BayesFlow} framework was nearly 7.5 hours, and the inference with MCMC was 0.9 hours per 10,000 posterior draws (Table 2). It also cost nearly 4.5 hours for MCMC to perform inference for a single dataset. In this case, the advantage of amortized inference is substantial as the extra effort of learning the global \textit{BayesFlow} model upfront would be worthwhile even when considering as few as two datasets.

In these selected datasets, in addition to being more computationally efficient than MCMC, based on learning standardized summary statistics from raw data, \textit{BayesFlow} also slightly outperformed MCMC in estimating the parameters $\rho$ and $\sigma^2$ for the 2-D LGCPs (Table 2). 
Similarly to its performance for all 300 simulated datasets in the 1-D LGCP scenario, \textit{BayesFlow} had a tendency to overestimate $\rho$ for 2-D point patterns in which the spatial dependency was weaker (Figure 3). 

\vspace{0.1in}
\begin{table}[H]
\centering
\begin{tabular}{c c c c c c c c} 
 \hline
  &  & \textit{BayesFlow} & \textit{BayesFlow}(standardize)   &  MCMC  \\ [0.5ex] 
\hline
NRSSE & $\mu$      & 7.139  & 4.104  & 3.619\\
      &  $\rho$    & 1.865  & 1.667  & 1.475 \\
      & $\sigma^2$ & 6.140  & 5.285  & 4.649\\
\hline      
$R^2$ &  $\mu$     & 0.308  & 0.771  & 0.822\\
      &  $\rho$    & 0.095  & 0.277  & 0.434\\
      & $\sigma^2$ & 0.284  & 0.470  & 0.589\\
\hline      
Upfront Training   & -      & 8.6 h  & 12.5 h          &  - \\
Inference (per dataset with 10,000 posterior samples)  & -  & 2.5 s  & 2.5 s &  8.4 s\\
\hline
\end{tabular}
\caption{Report performance metrics in terms of NRSSE and R$^2$ on 300 different simulated 1-D LGCP datasets for each parameter and each method (two \textit{BayesFlow} methods and MCMC). The smaller NRSSE, the better point-estimates across all validation test datasets.}
\label{table:1}
\end{table}

\begin{table}
\centering
\begin{tabular}{c ||c|c|c|c} 
 \hline
 & & Truth & \textit{BayesFlow} (Mean, CI)   &  MCMC (Mean, CI)  \\ [0.5ex] 
\hline
Simulated Data 1 & $\mu$       & 5.447   & 5.058 [3.531, 5.921]  & 5.656 [5.169, 5.980]\\
                 & $\rho$      & 0.106   & 0.094 [0.043, 0.133]  & 0.130 [0.102, 0.149] \\
                 & $\sigma^2$  & 1.758   & 1.574 [0.835, 1.938]  & 1.731 [1.345, 1.984]\\
\hline
Simulated Data 2 & $\mu$       & 5.510   & 5.171 [3.462, 5.968]  & 5.811 [5.494, 5.995]\\
                 & $\rho$      & 0.102   & 0.115 [0.076, 0.139]  & 0.103 [0.081, 0.129]\\
                 & $\sigma^2$  & 1.850   & 1.760 [1.304, 1.959]  & 1.659 [1.286, 1.976]\\
\hline
Simulated Data 3 & $\mu$       & 5.800   & 5.227 [3.655, 5.948]  & 5.804 [5.387, 5.992] \\
                 & $\rho$      & 0.110   & 0.101 [0.052, 0.136]  & 0.129 [0.101, 0.149]\\
                 & $\sigma^2$  & 1.717   & 1.560 [0.806, 1.935]  & 1.437 [1.136, 1.709]\\
\hline
NRSSE & $\mu$      & -  & 1.298           & \textbf{0.617} \\
      &  $\rho$    & -  & \textbf{0.217}  & 0.348 \\
      & $\sigma^2$ & -  & \textbf{0.709}  & 0.933 \\
\hline            
Upfront Training   & -  & -    & 7.5 h &  - \\
Inference (per dataset with 10,000 posterior samples)  & -  & - & 2.1 s &  53.4 min\\
\hline
\end{tabular}
\caption{Report detailed posterior results (posterior mean and 95$\%$ credible intervals) and performance metric (NRSSE) on selective simulated 2-D LGCP datasets for each parameter from \textit{BayesFlow} (with standardized summary statistics as conditional input) and MCMC. MCMC is fit based on uniform $50 \times 50$ grids resolution. For metric associated with each parameter, the best performance across methods is printed in bold font.}
\label{table:2}
\end{table}

\begin{figure}
 \centering
  \includegraphics[width=0.32\linewidth]{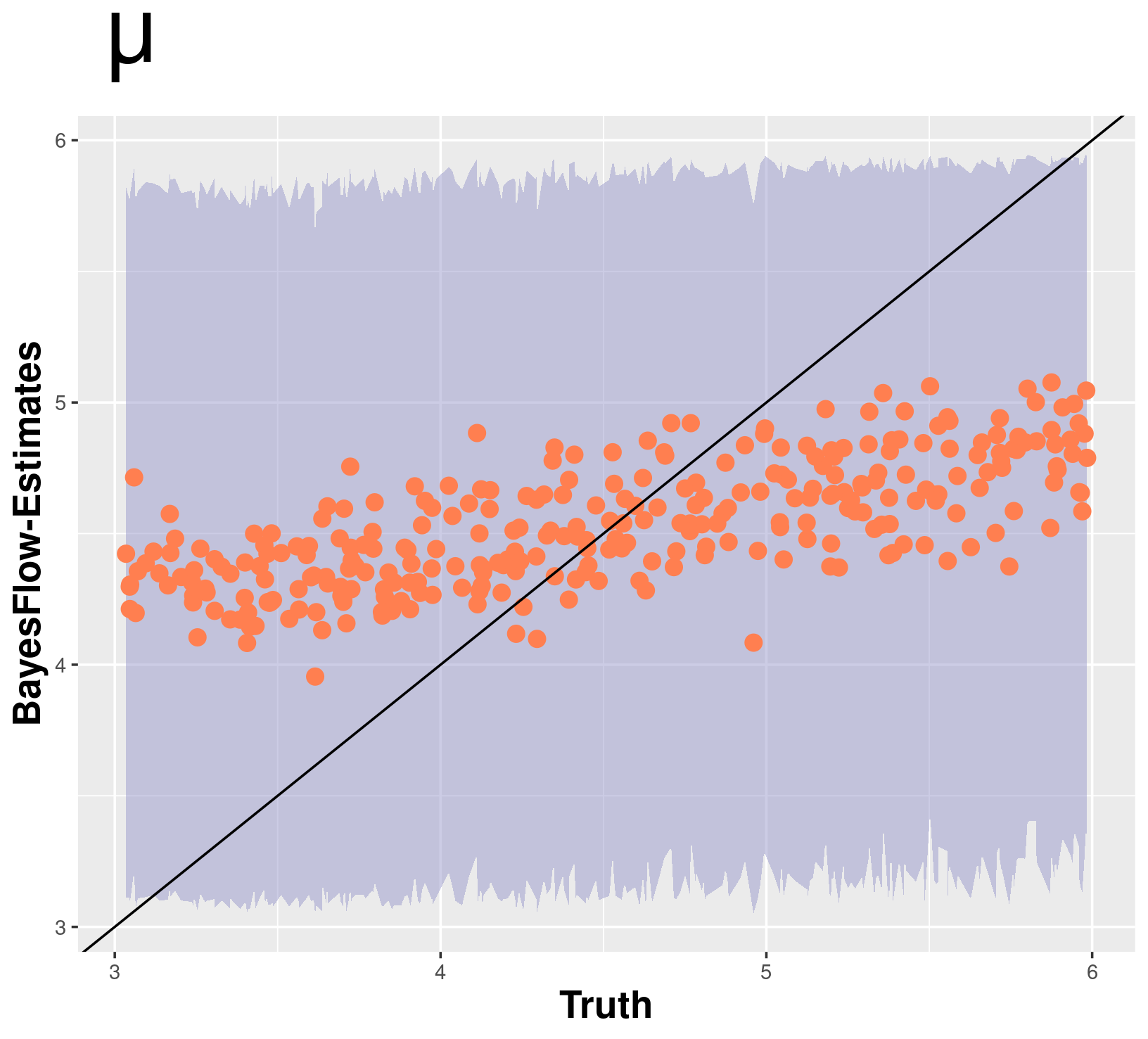}
  \includegraphics[width=0.32\linewidth]{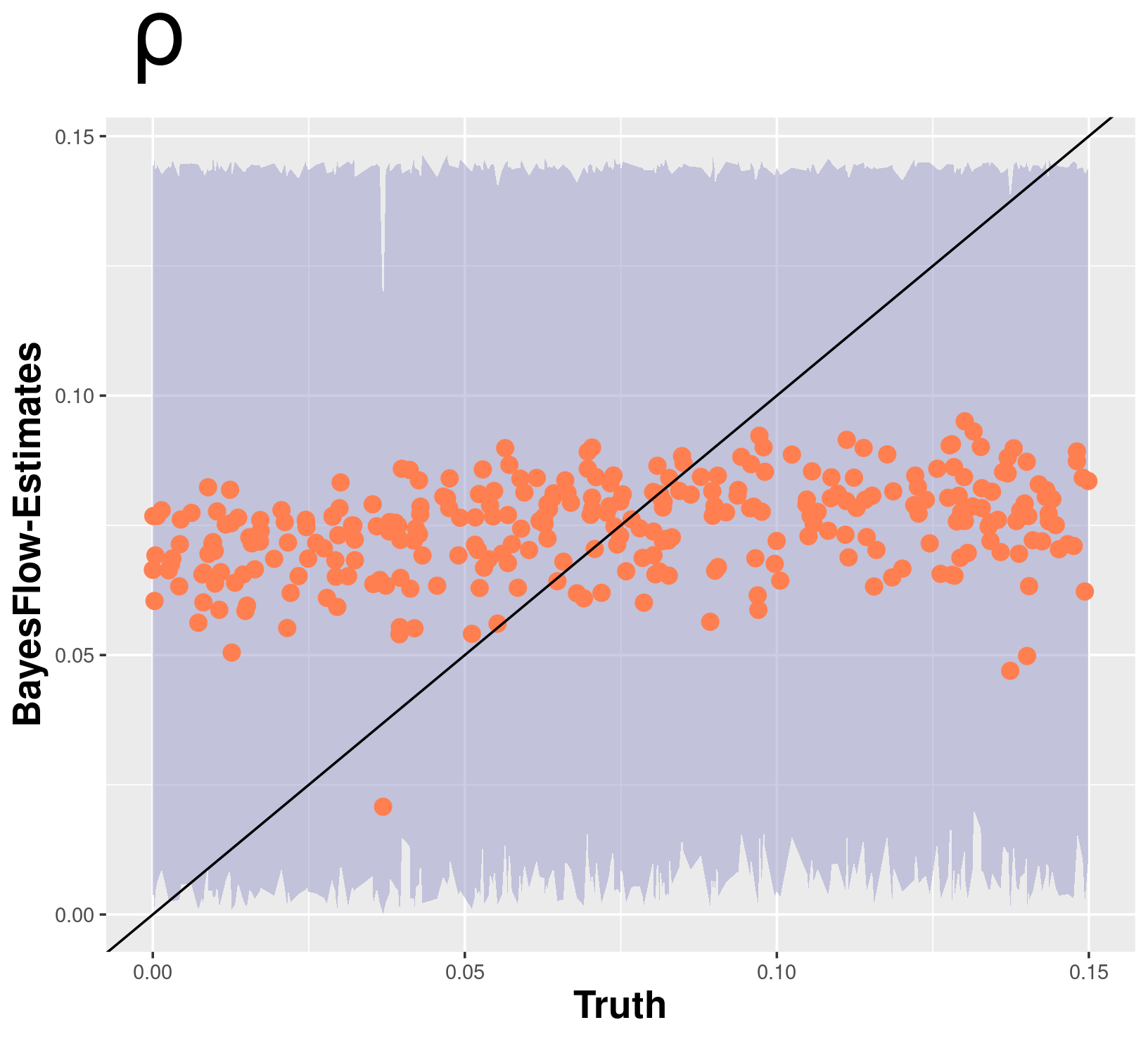}
  \includegraphics[width=0.3208\linewidth]{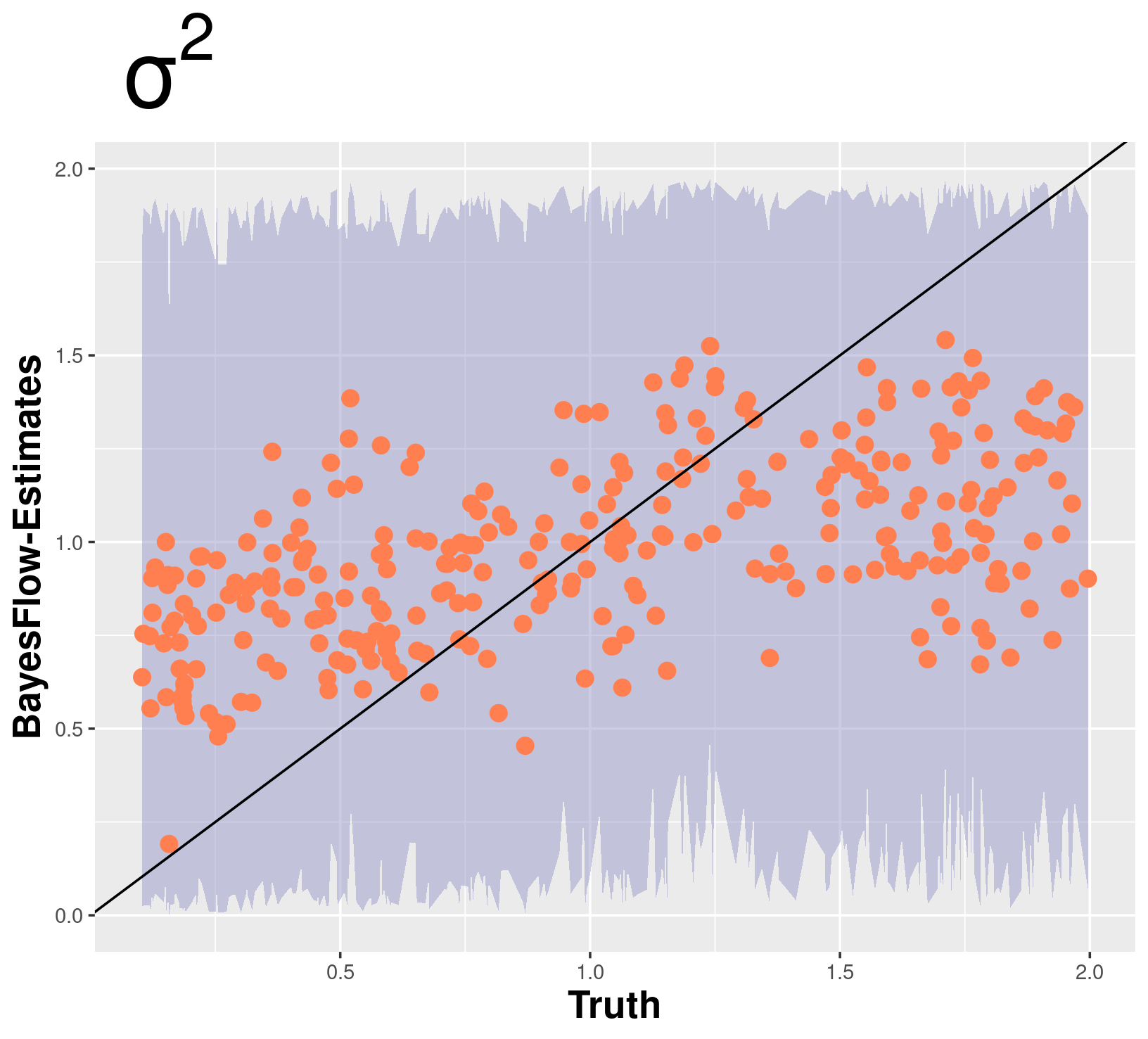}
  \includegraphics[width=0.32\linewidth]{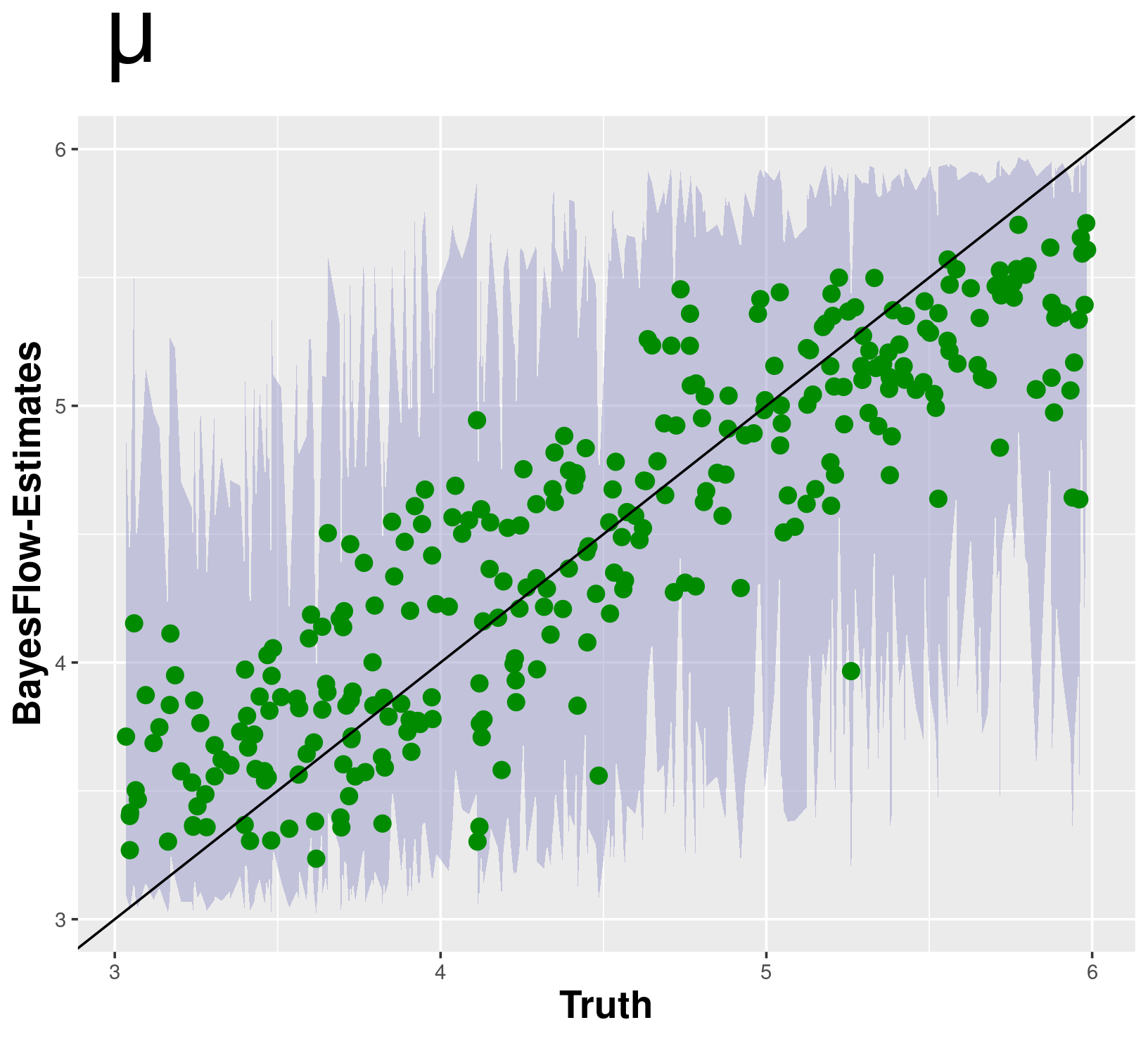}
  \includegraphics[width=0.32\linewidth]{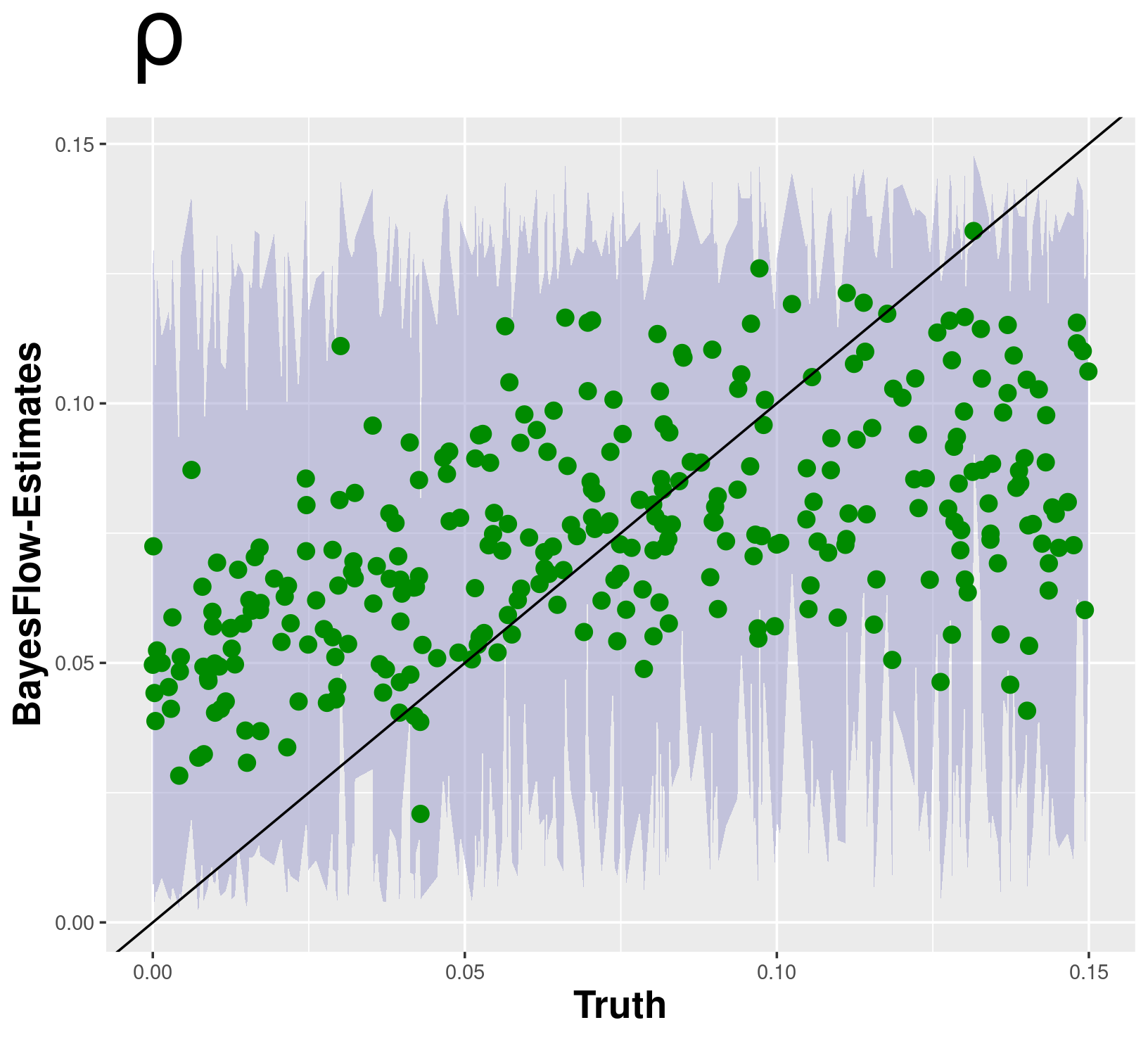}
  \includegraphics[width=0.3208\linewidth]{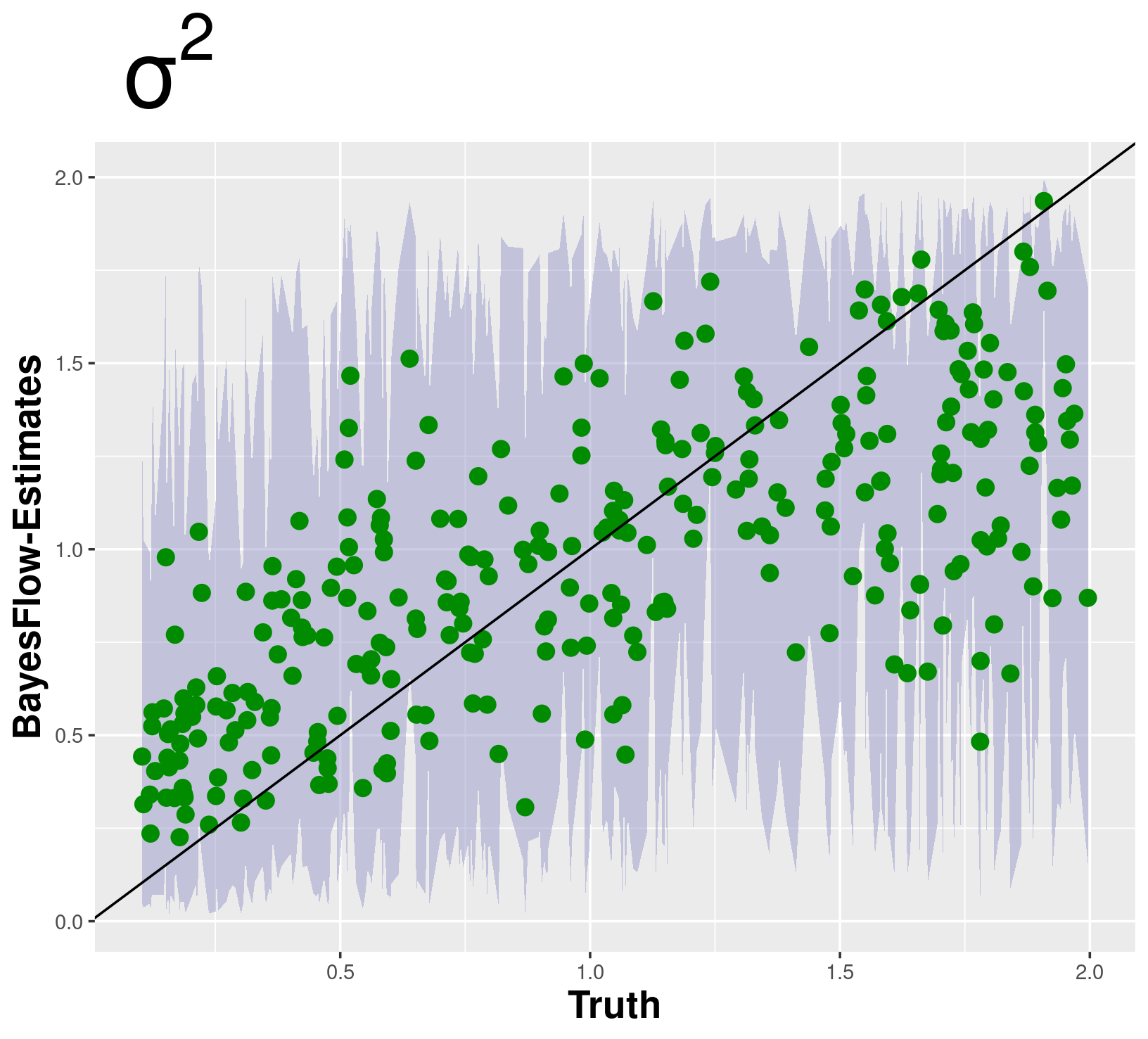}
  \includegraphics[width=0.32\linewidth]{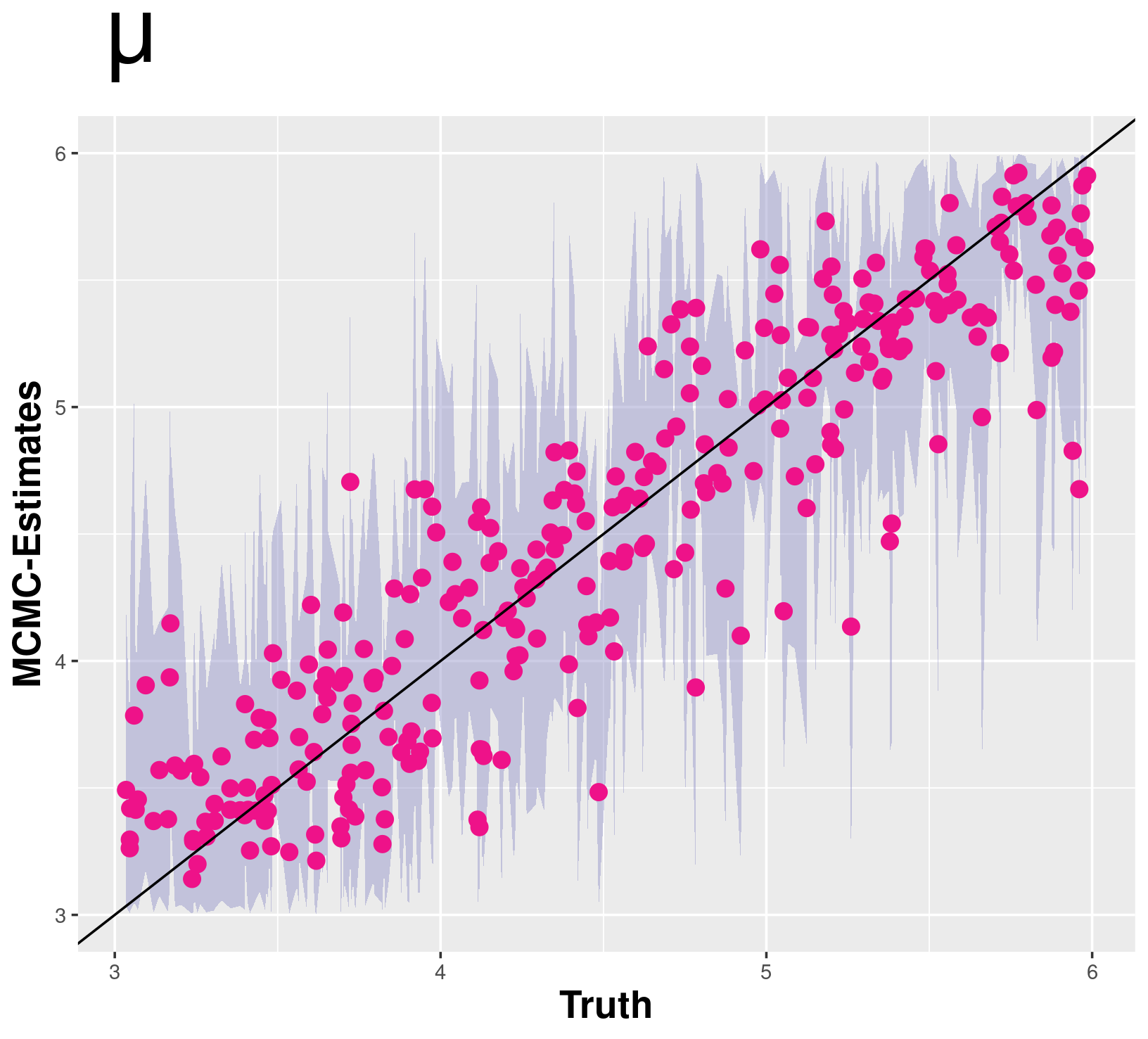}
  \includegraphics[width=0.32\linewidth]{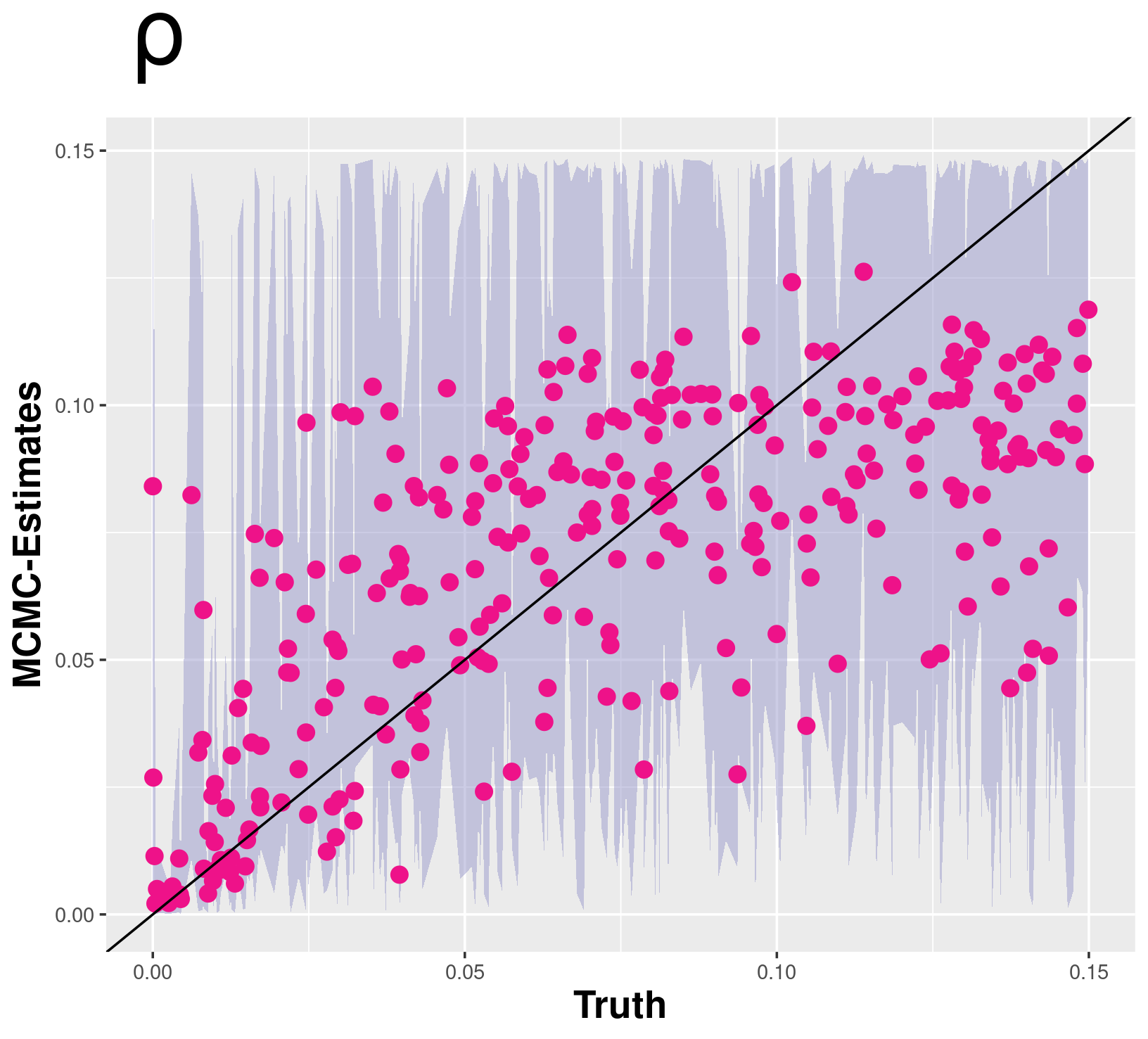}
  \includegraphics[width=0.3208\linewidth]{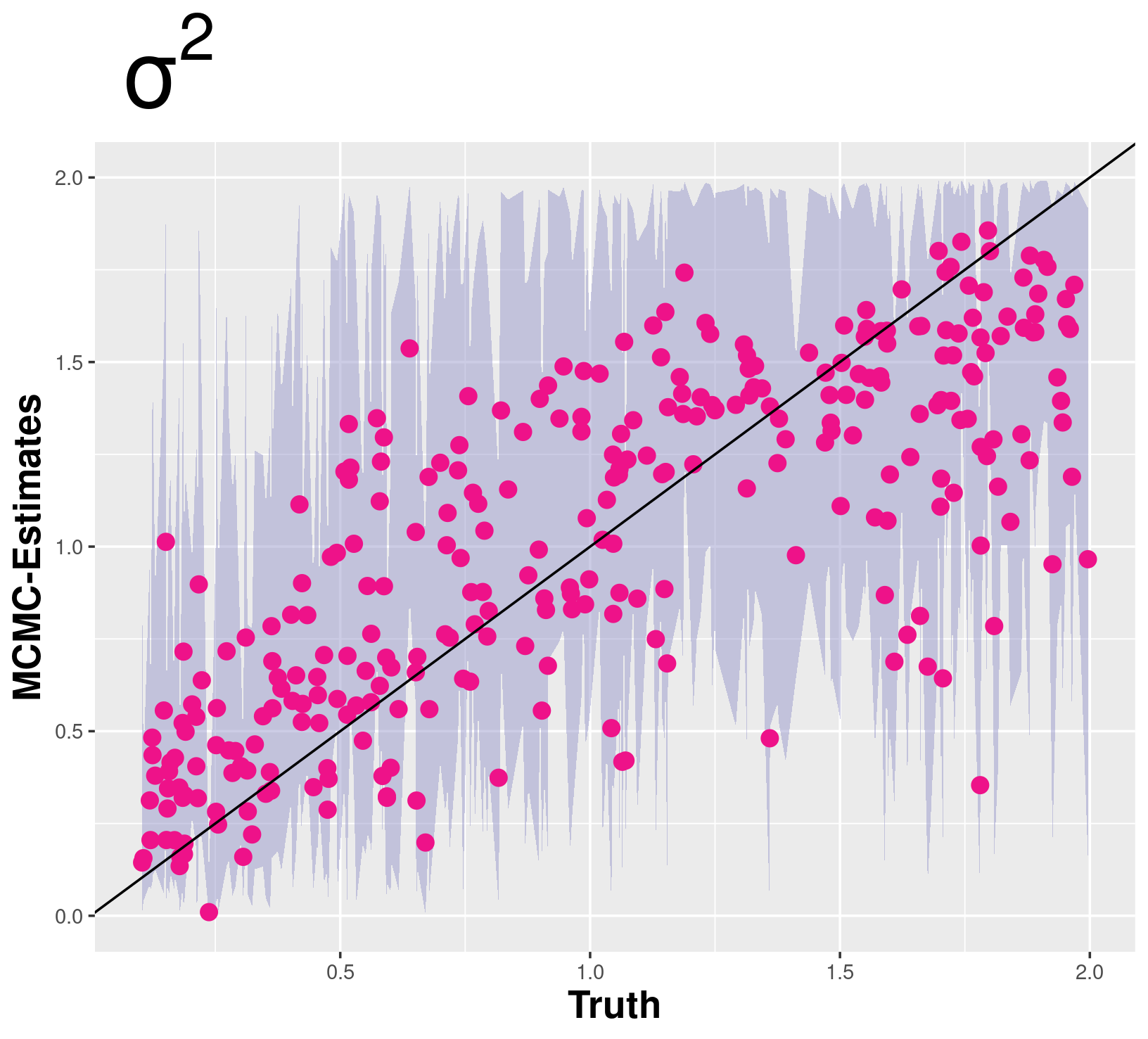}
  \caption{Parameter recovery comparison between \textit{BayesFlow} and MCMC on 300 different simulated 1-D LGCP data sets of true parameters. The first row are \textit{BayesFlow} mean estimates versus the truth for each parameter using original summary statistics as conditional input; the second row are \textit{BayesFlow} mean estimates versus the truth for each parameter using standardized summary statistics as conditional input; the third row are MCMC posterior mean estimates versus the truth for each parameter. The black solid line represent the identity line for comparison reference. The shaded region shows posterior 95$\%$ credible intervals for each parameter.}
\end{figure}

\begin{figure}
 \centering
  \includegraphics[width=0.320\linewidth]{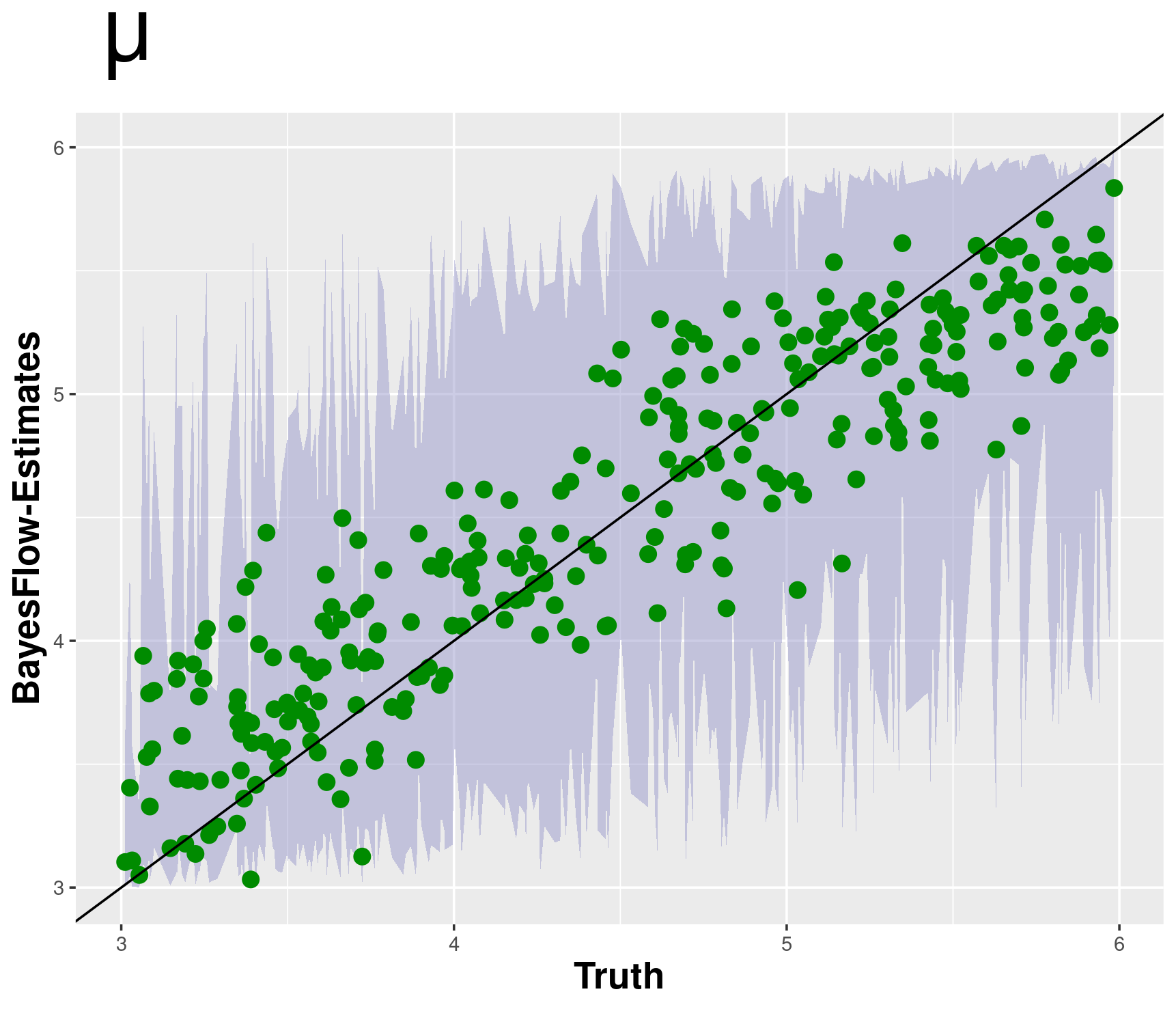}
  \includegraphics[width=0.320\linewidth]{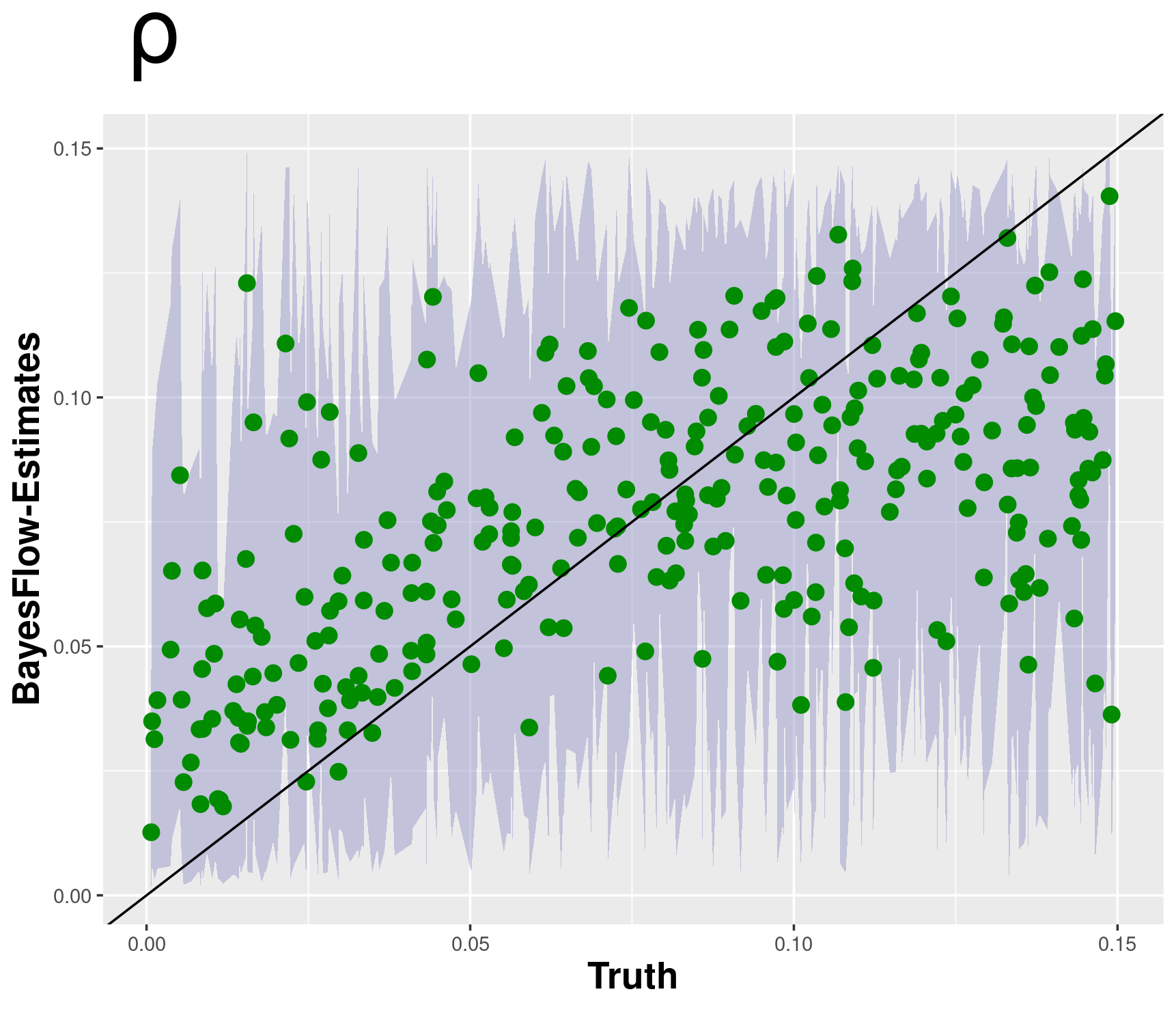}  
  \includegraphics[width=0.329\linewidth]{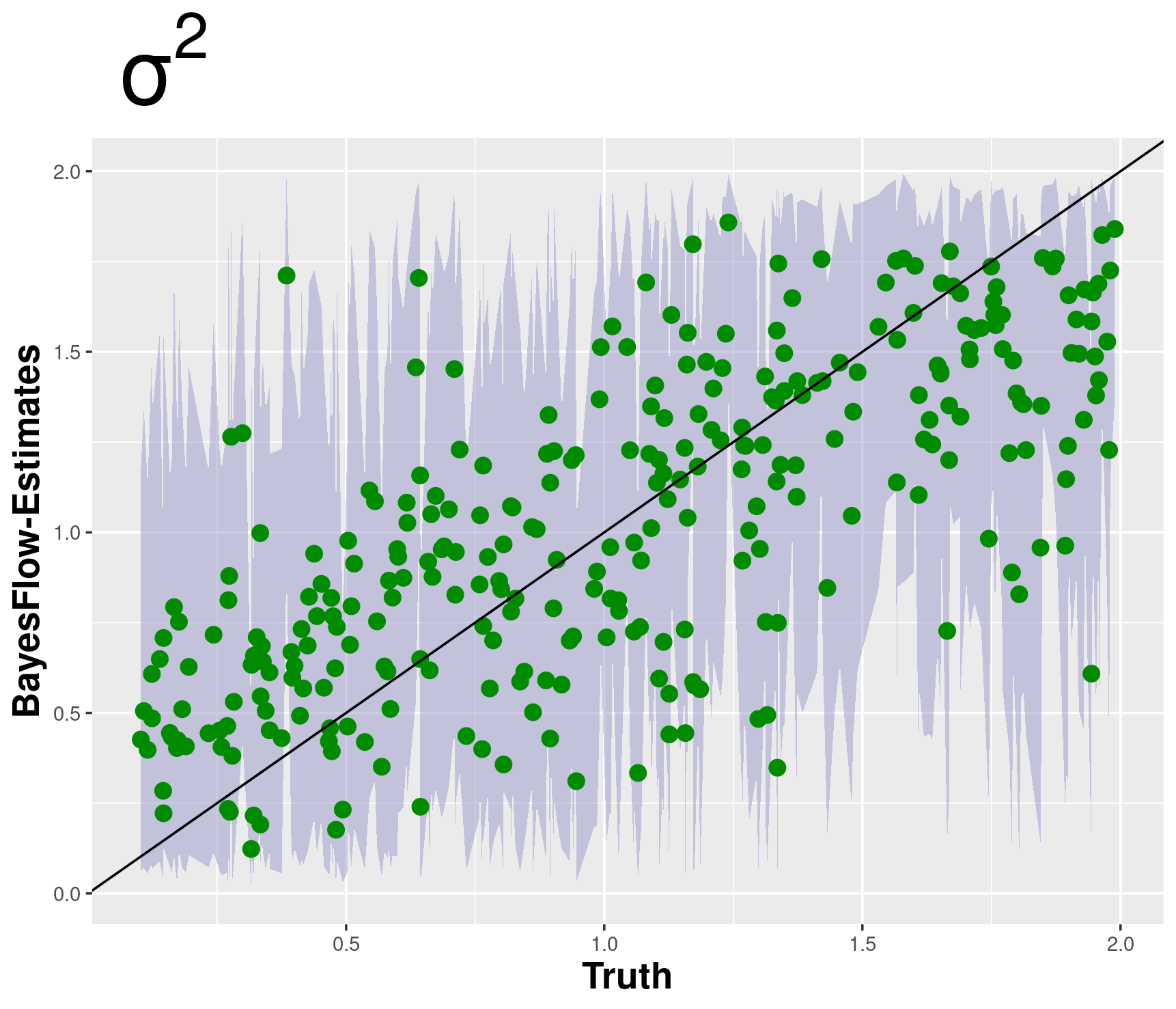}
  \caption{Parameter recovery performance from \textit{BayesFlow} on 300 different simulated 2-D LGCP data sets of true parameters. The y-axis are \textit{BayesFlow} mean estimates and the x-axis are the truth for each parameter. The black solid line represent the identity line for comparison reference. The shaded region shows posterior 95$\%$ credible intervals for each parameter.}
\end{figure}

\section{Application}

We demonstrate the practical utility of the proposed framework by applying it to oral microbial biofilm image data. Specifically, we used the proposed method to investigate the spatial point pattern structures of five bacterial types (taxa) within the images (Figure 4).  This application illustrates the computational benefits of our method. Specifically, the conventional MCMC approach requires performing the inference procedure repeated from the beginning for each individual point pattern. In contrast, as shown in Section 3, the Bayesian neural network inference method makes use of ``amortization", which is computationally efficient for estimating parameters associated with each of the multi-type point patterns observed within the same domain, after a one-time training procedure. We describe the collection of biofilm image data in Section 4.1, the process of extracting spatial coordinates in Section 4.2, and show the inference performance and validation in Section 4.3.

\textbf{4.1 Microbial Biofilm Image Data from Human Tongue Dorsum}

Detailed data collection methods for microbiome biofilm image data were described in \cite{wilbert2020spatial}. Briefly, the microbial biofilm from the human tongue dorsum was collected by scraping a ridged plastic tongue scraper over the tongue from back to front. 
The biofilm was fixed in ethanol to preserve its spatial structure, then was gently spread onto slides for multiplexed fluorescence spectral imaging of microbial consortia (complexes of bacterial cells) ranging from tens to hundreds of microns in linear dimension. \par
The biofilm images revealed a highly structured spatial organization of microbes, with individual taxa forming single-taxon or multi-taxon patches. To identify the distribution of major microbes, multiplexed, spectral imaging fluorescence \textit{in situ} hybridization (FISH) was performed targeting the 17 abundant genera as well as 7 abundant species within these genera, plus 1 phylum. A total of 100 original images (20 images per subject across 5 subjects) were generated, and the following taxa presented in all subjects and in $\geq 69 \%$ of images acquired (supplemental Table 4 in \citealp{wilbert2020spatial}): \textit{Rothia}, \textit{Actinomyces}, \textit{Streptococcus}, \textit{Veillonella}, and \textit{Neisseriaceae} (genus \textit{Actinomyces} has recently been split into two genera,
\textit{Actinomyces} and \textit{Schaalia}. Both were abundant on the tongue dorsum and for simplicity we hereafter referred to both as \textit{Actinomyces}). We investigated the spatial distributions of three genera (\textit{Rothia}, \textit{Actinomyces}, and \textit{Veillonella}) and two species (\textit{Streptococcus mitis} and \textit{Streptococcus salivarius}), excluding \textit{Neisseriaceae} due to its relative limited/rare observance on the tongue dorsum.   

\textbf{4.2 Image Processing}

We focused analyses on two biofilm images that had distinct domain shapes, i.e., different shapes of the area in which any bacterial cells could be found (Figure 4). Among the material scraped from the tongue, including sparsely colonized epithelial cells and loose bacteria,  consortia were defined as objects with a well-defined perimeter and a core composed of host epithelial cells. Images were segmented to separate cells from background and the spatial coordinates for each cell's centroid were generated using $\textit{Fiji}$ \citep{schindelin2012fiji} with the bottom-left corner of the image assigned the coordinate $(0,0)$ after removing unnecessary regions without any observations. There were five abundant taxa present in these two tongue biofilm consortia (rescaled to unit planar window $[0,1]^2$), clusters of \textit{Rothia}, \textit{Actinomyces}, \textit{S. mitis} and \textit{S. salivarius} and \textit{Veillonella}, respectively (Figures 5 $\&$ 6).

\begin{figure}[H]
\centering
     \begin{subfigure}[b]{0.4\textwidth}
         \includegraphics[width=\textwidth]{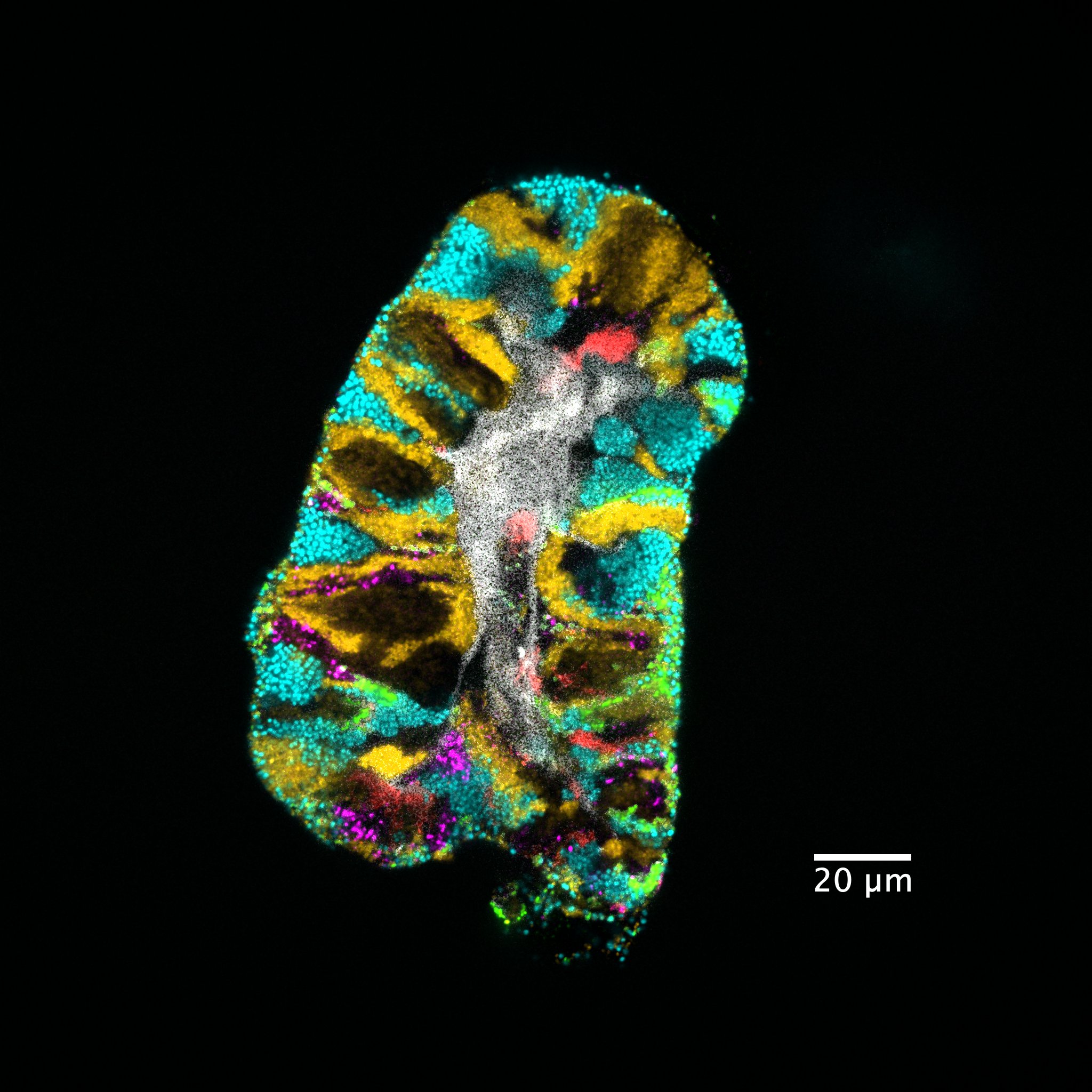}
         \caption{}
     \end{subfigure}
     \hspace{0.2in}
     \begin{subfigure}[b]{0.4\textwidth}
         \includegraphics[width=\textwidth]{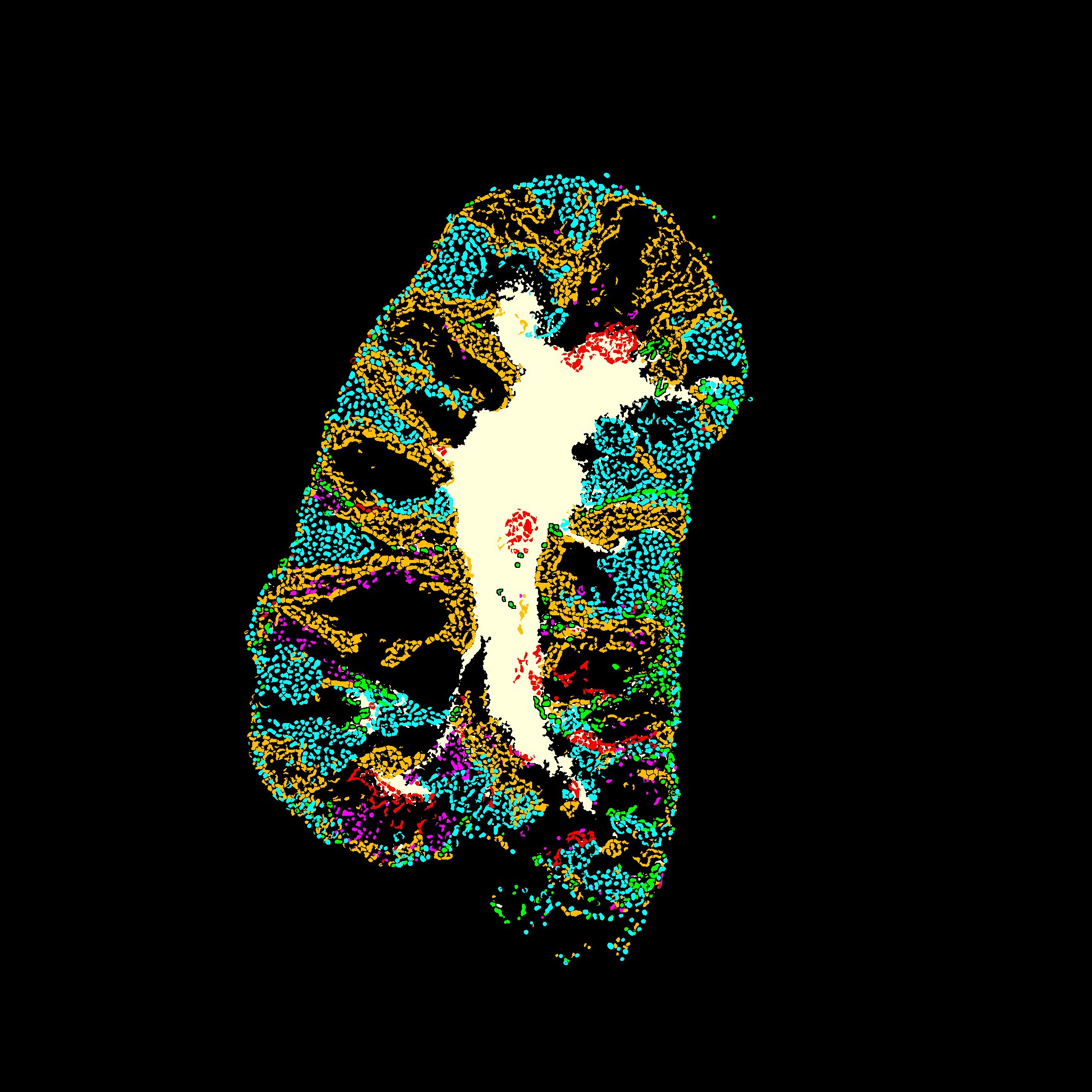}
         \caption{}
     \end{subfigure}
     \hfill
     \\
     \centering
     \begin{subfigure}[b]{0.4\textwidth}
         \includegraphics[width=\textwidth]{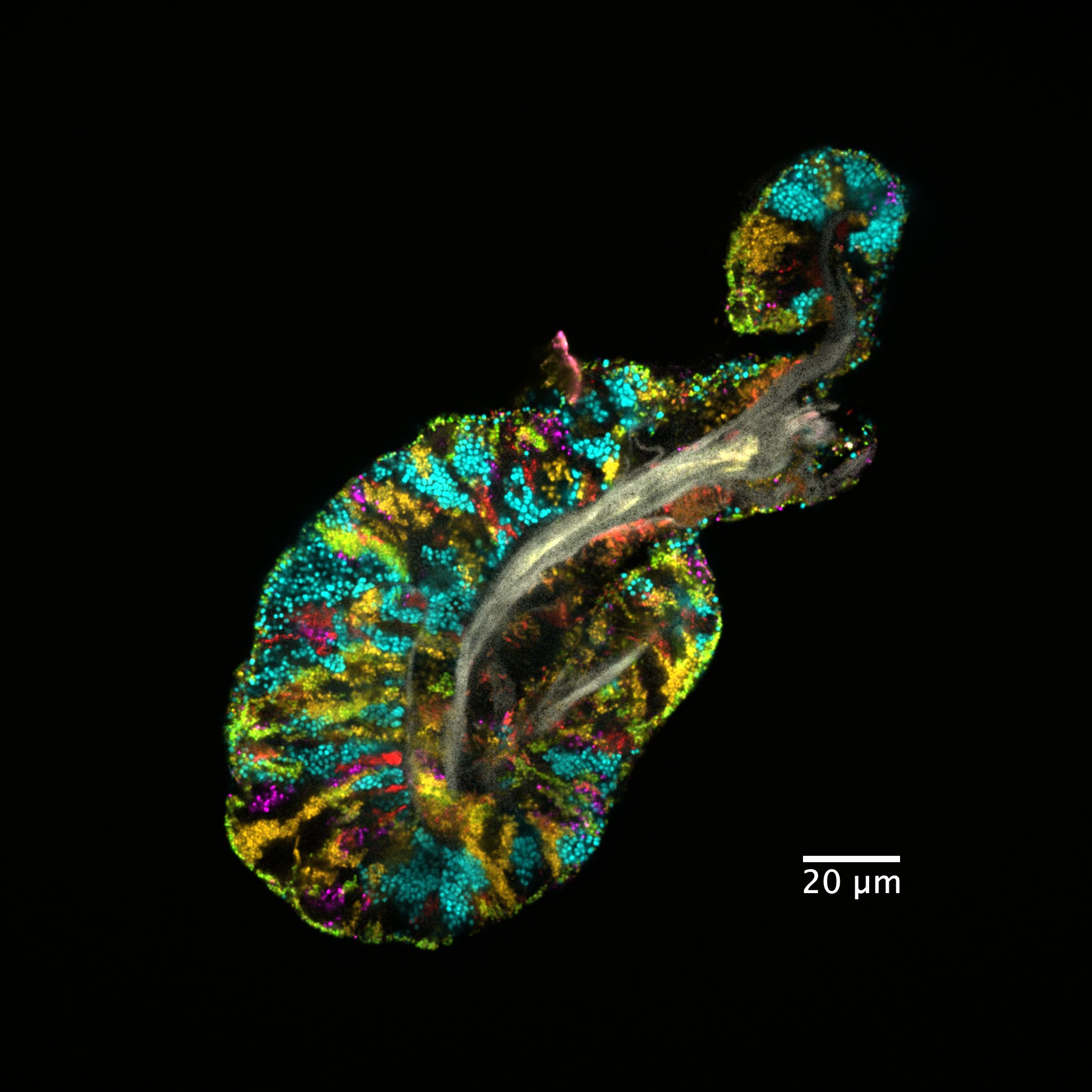}
         \caption{}
     \end{subfigure}
     \hspace{0.2in}
     \begin{subfigure}[b]{0.4\textwidth}
         \includegraphics[width=\textwidth]{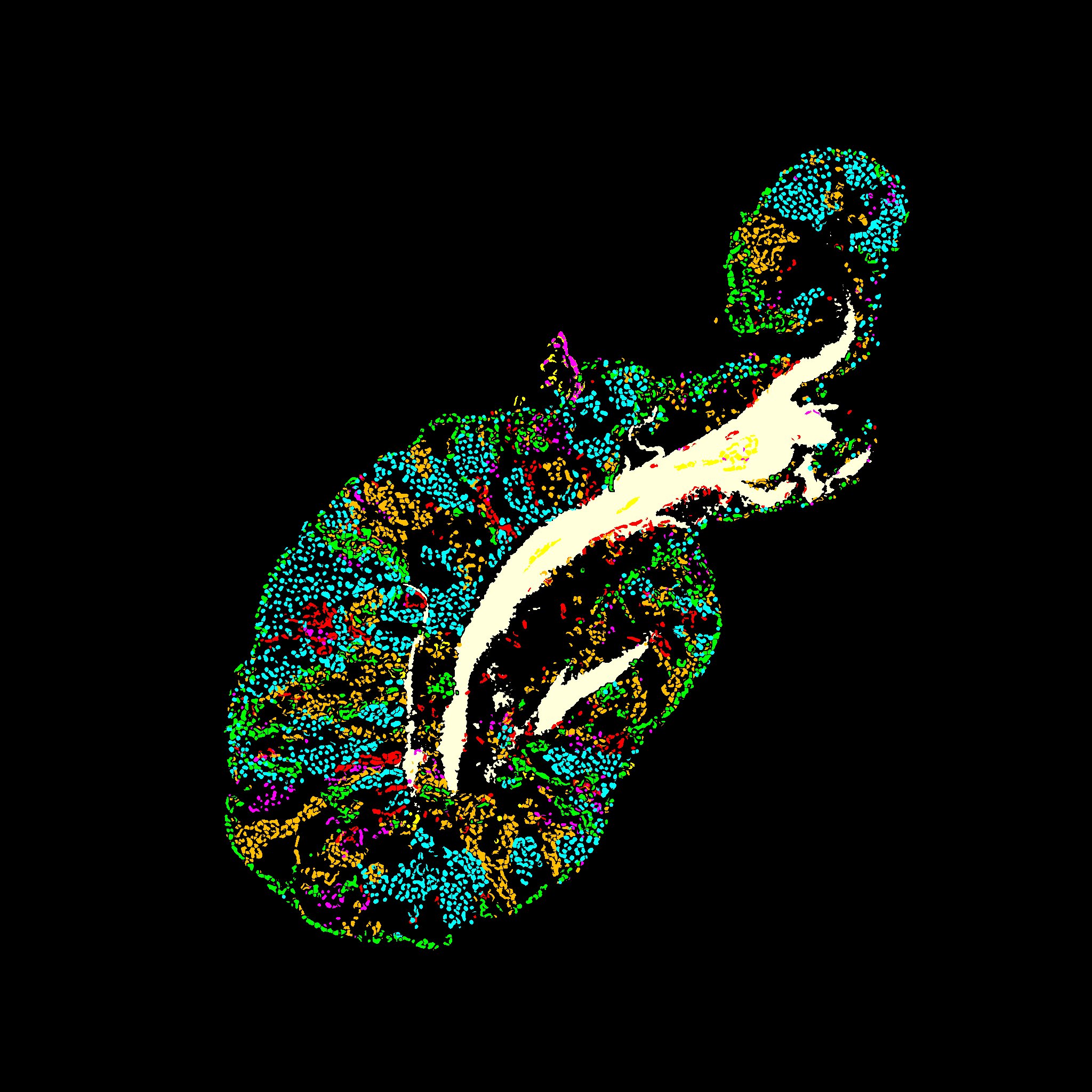}
         \caption{}
     \end{subfigure}
     \hfill
        \caption{Two images of the human tongue dorsum biofilm collected from different subjects (image A and image B) are shown in two versions, respectively: (a) $\&$ (c) unsegmented and (b) $\&$ (d) segmented. These images illustrate the spatial distribution of five bacterial taxa: \textit{Rothia} (Cyan), \textit{Actinomyces} (Red), \textit{Veillonella} (Magenta), \textit{S. mitis} (Green) and \textit{S. salivarius} (Orange). Host epithelial cells, identified by autofluorescence, are displayed in white. }
\end{figure}

\begin{figure}[H]
 \centering
  \includegraphics[width=0.95\linewidth]{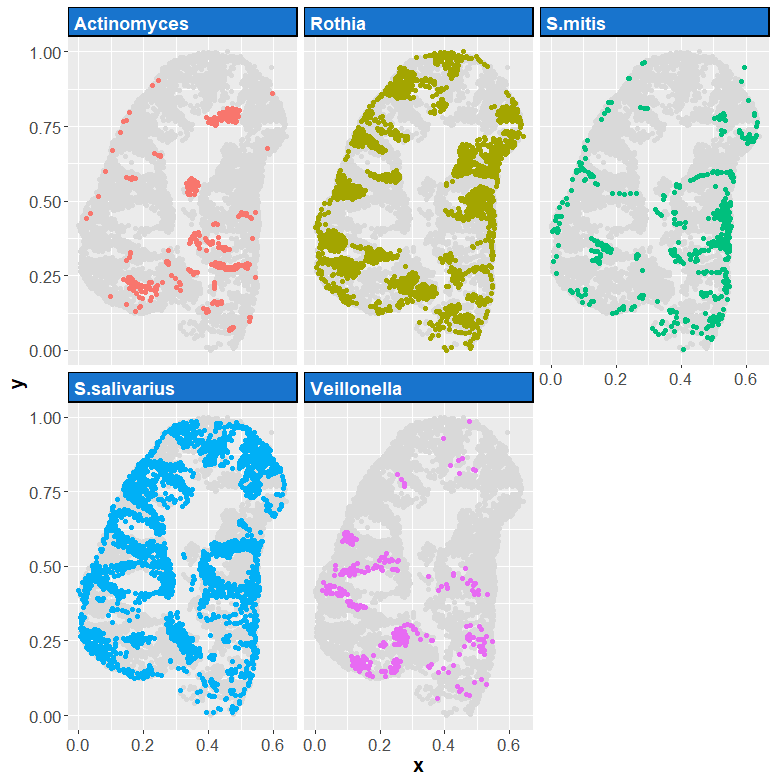}
  \caption{The spatial structure of each of the five taxa in the microbial consortium from tongue biofilm image A. Clusters of \textit{Rothia} (1386 observations), \textit{Actinomyces} (267 observations), \textit{S. mitis} (381 observations) and \textit{S. salivarius} (1847 observations) and \textit{Veillonella} (277 observations) are presented, respectively.}
\end{figure}

\begin{figure}[H]
 \centering
  \includegraphics[width=0.95\linewidth]{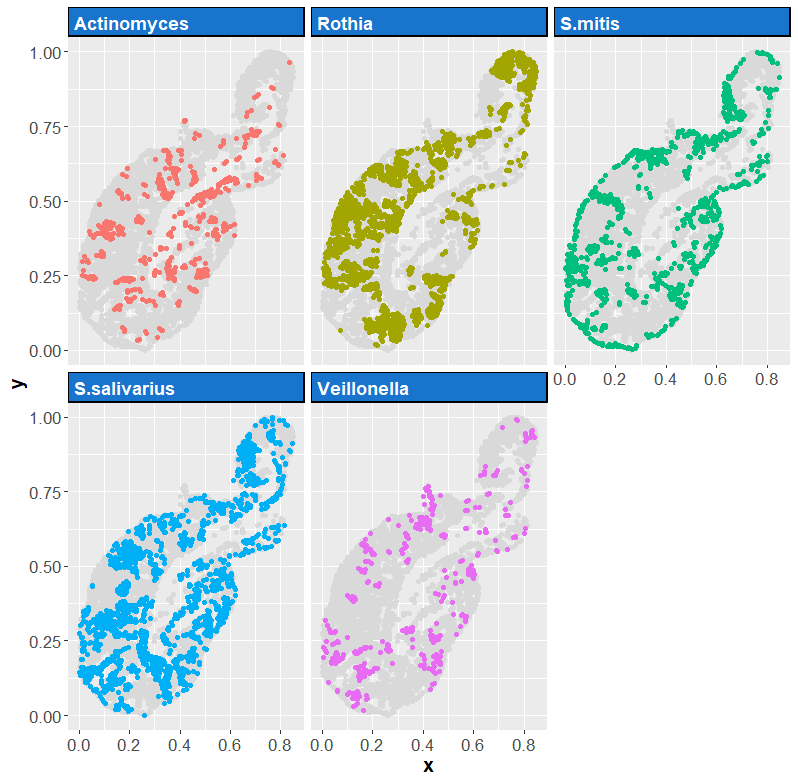}
  \caption{The spatial structure of each of the five taxa in the microbial consortium from tongue biofilm image B. Clusters of \textit{Rothia} (1342 observations), \textit{Actinomyces} (489 observations), \textit{S. mitis} (803 observations) and \textit{S. salivarius} (1418 observations) and \textit{Veillonella} (374 observations) are presented, respectively.}
\end{figure}
\textbf{4.3 Analyses}

Both the black regions outside microbial consortia and the host epithelial cells (white regions in Figure 4b $\&$ 4d) were excluded from the analysis of each image as is now considered standard to minimize bias (e.g., \citealp{schillinger2012co}). Accordingly, both regions were excluded from the spatial domain using indicator variables on the computational grid when simulating data sets for training phase in \textit{BayesFlow}, i.e., the random intensity function of a LGCP in Eq.(1) becomes
\begin{align}
\lambda(\boldsymbol{s})=exp(\boldsymbol{Z(s)}) \times I(\boldsymbol{s} \in \textit{W}_r ), \hspace{0.2in} \tag{10}
\end{align}
where $ \textit{W}_r $ stands for the non-excluded grid regions in the bounded window. \par
We then used the converged network parameters $\hat{\boldsymbol{\phi}}$ from the upfront training to make posterior inference for the parameters of interest that reflect spatial self-association (clustering with self) for three genera (\textit{Rothia}, \textit{Actinomyces}, and \textit{Veillonella}) and two species (\textit{Streptococcus mitis} and \textit{Streptococcus salivarius}). To validate the obtained estimates, we considered point-wise zero-probability function \citep{micheas2018theory} envelope plots based on posterior predictions using posterior mean estimators from \textit{BayesFlow} for each taxon. \par
Specifically, given the estimated parameter vector $\boldsymbol{\theta}= (\mu, \rho, \sigma^2)'$ from \textit{BayesFlow}, 10,000 realizations were generated on the specified domain and used to construct point-wise zero-probability function envelope plots. The zero-probability function can be interpreted as the probability that a point pattern misses a closed test set (e.g., a ball centered at the origin with fixed radius), and it uniquely determines the probability distribution of the point pattern (see more details in \citealp{micheas2018theory}). The zero-probability function estimates from the observed point patterns (red dotted line) lay within 95$\%$ simulated envelopes of the zero-probability function estimates (Figures 7 $\&$ 8). These results suggest that the proposed LGCP model was appropriate to describe the spatial self-association of each bacterial taxon. Among the five taxa, the observed functional statistics of \textit{Actinomyces} and \textit{Veillonella} (red dotted line) deviated more from simulated mean estimates (blue dashed line), possibly due to relative spatial sparseness. \par
To test whether the biomass domain mattered for the evaluation of the spatial distribution, we also conducted the same validation procedure for the estimates obtained using the converged network parameters $\hat{\boldsymbol{\phi}}$ from the 2-D LGCP simulation studies (Figure 9). Recall, in the 2-D LGCP simulation studies, the training phase was based on realizations in the whole window $\textit{W}$ without any exclusions. We evaluated such obtained estimates for \textit{Veillonella} in Figure 4(b) and \textit{Streptococcus mitis} in Figure 4(d). The zero-probability function estimates based on the observations (Figure 9, left) were not within the simulated interval, indicating that biased estimates were obtained, which agrees with the findings from previous studies that biases can be introduced without distinguishing biomass-containing regions from the background (e.g., \citealp{schillinger2012co}). For the \textit{Streptococcus mitis} (Figure 9, right), the red dotted line deviated from the simulated mean estimates (blue dashed line) much more than in Figure 8(d), indicating the obtained estimates were not as reasonable as when we accounted for the biomass-containing regions in microbial consortia. 

\begin{figure}
     \centering
     \begin{subfigure}[b]{0.32\textwidth}
         \includegraphics[width=\textwidth]{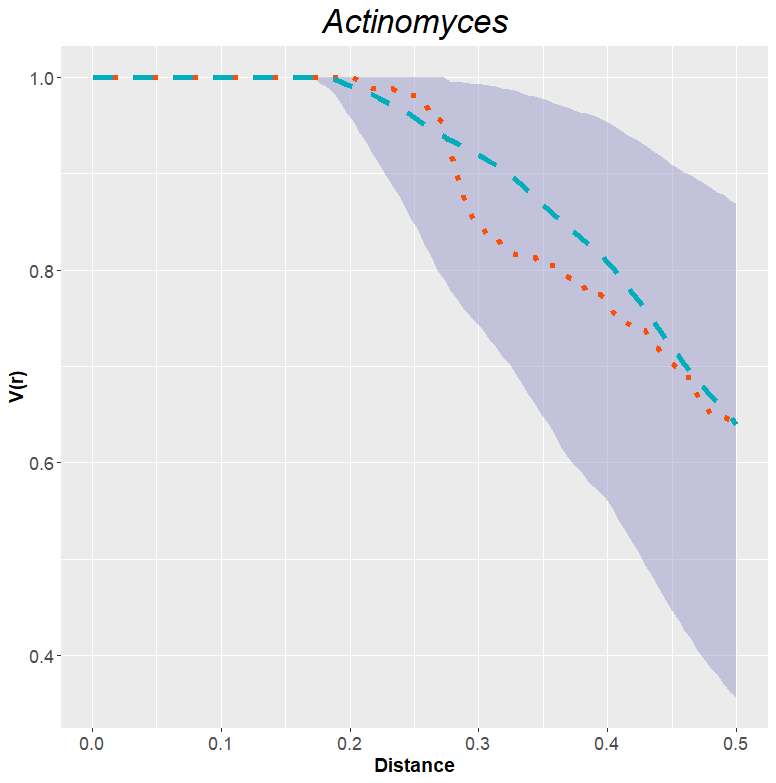}
         \caption{}
     \end{subfigure}
     \hfill
     \begin{subfigure}[b]{0.32\textwidth}
         \includegraphics[width=\textwidth]{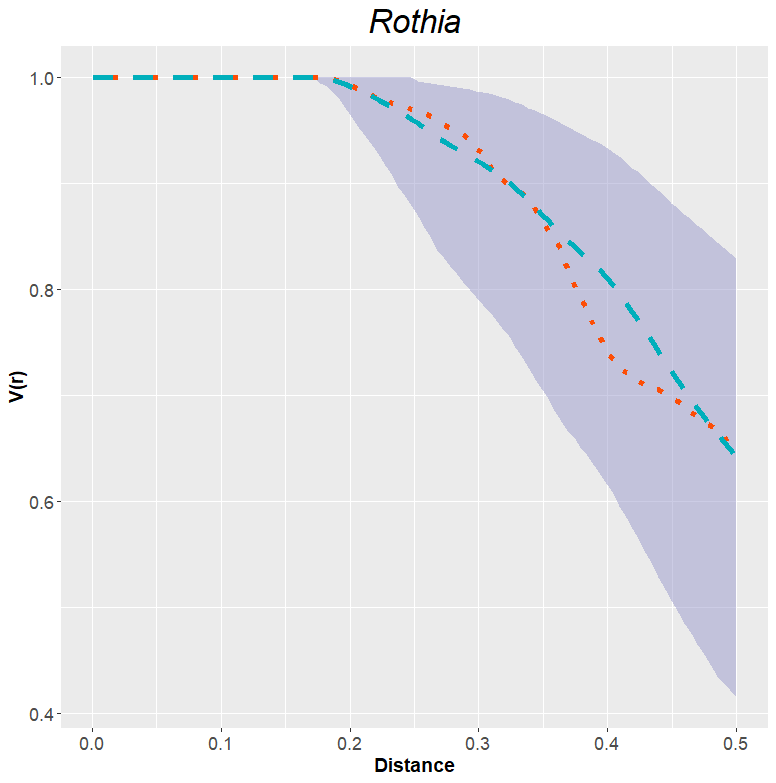}
         \caption{}
     \end{subfigure}
     \hfill
     \begin{subfigure}[b]{0.32\textwidth}
         \includegraphics[width=\textwidth]{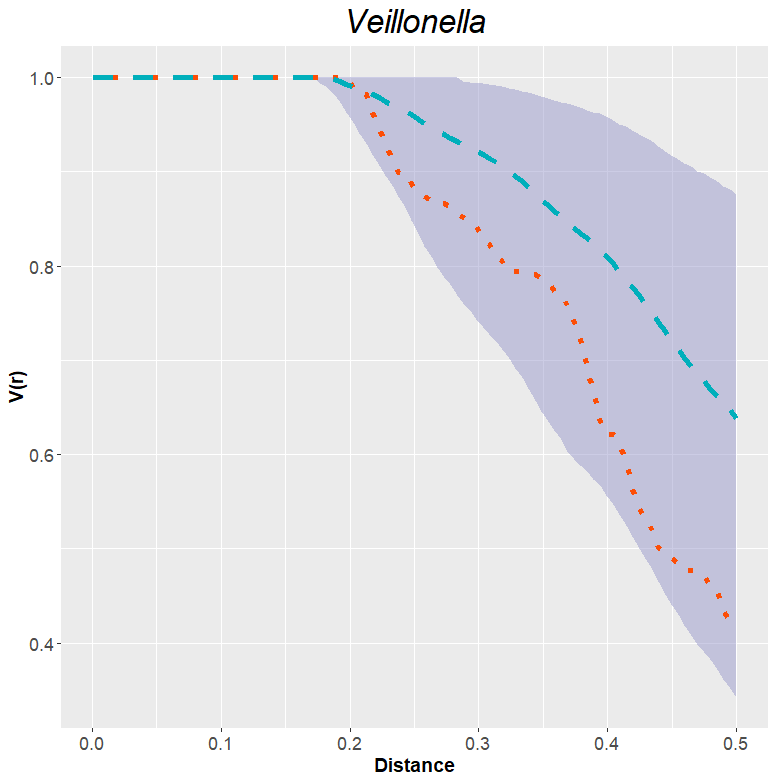}
         \caption{}
     \end{subfigure}
     \hfill
     \begin{subfigure}[b]{0.32\textwidth}
         \includegraphics[width=\textwidth]{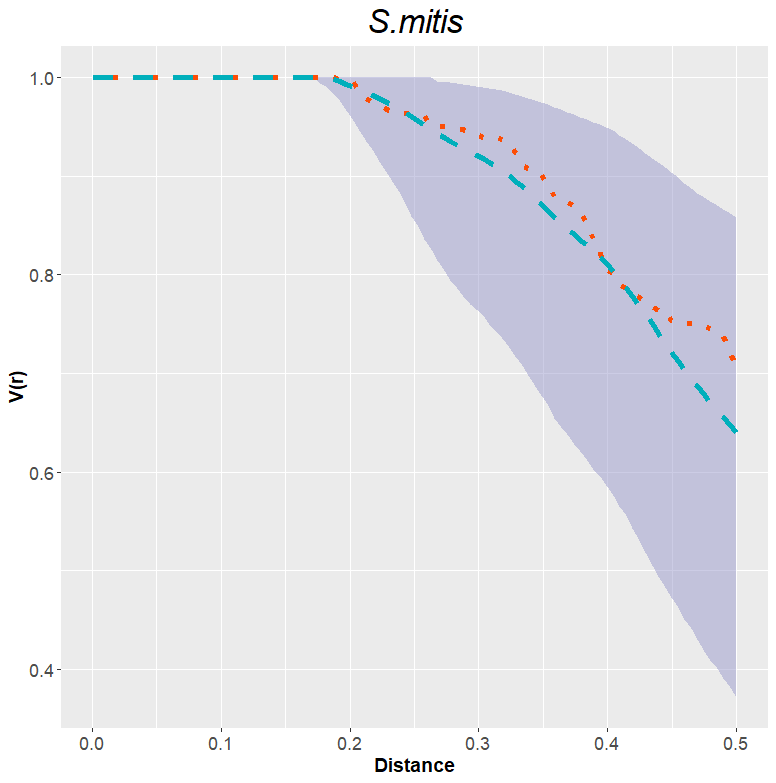}
         \caption{}
     \end{subfigure}
     \begin{subfigure}[b]{0.32\textwidth}
         \includegraphics[width=\textwidth]{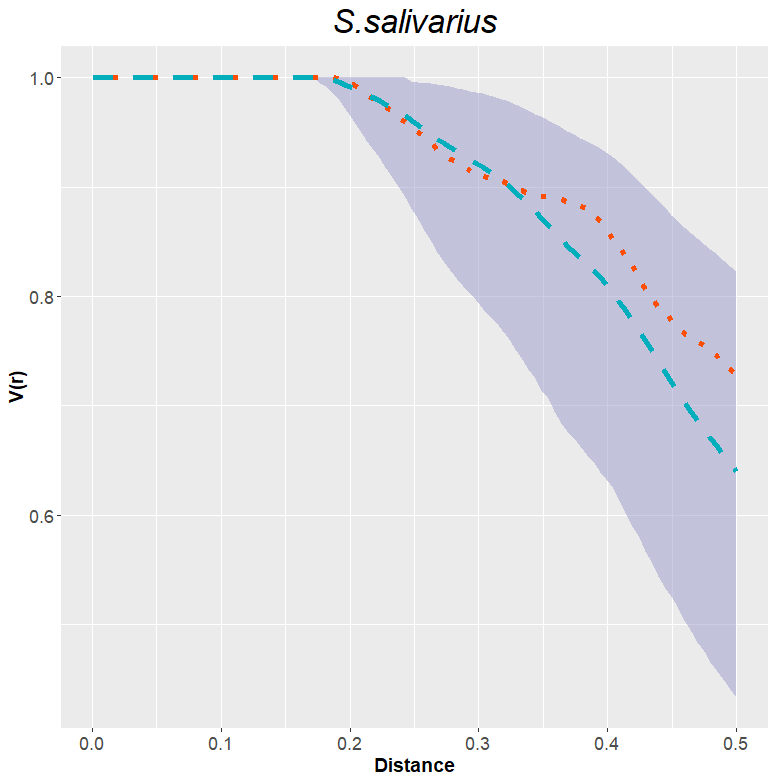}
         \caption{}
     \end{subfigure}
     \hfill
        \caption{Zero-probability function envelope plots for assessing posterior predictive distribution and model validation for each isolated taxa on the preserved tongue consortia from microbiome biofilm image A. 
        The function estimates from the observations (red dotted curve), the point-wise means (blue dashed curve) and 95$\%$ interval (shaded area) obtained from 10,000 simulated posterior predictions were summarized for each isolated taxa.}
\end{figure}

\begin{figure}
     \centering
     \begin{subfigure}[b]{0.32\textwidth}
         \includegraphics[width=\textwidth]{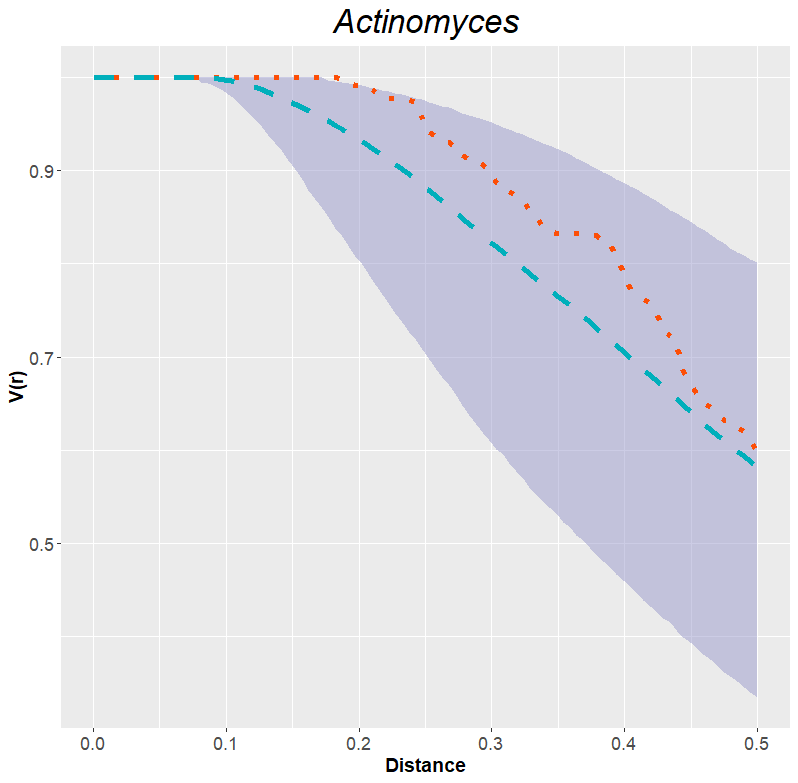}
         \caption{}
     \end{subfigure}
     \hfill
     \begin{subfigure}[b]{0.32\textwidth}
         \includegraphics[width=\textwidth]{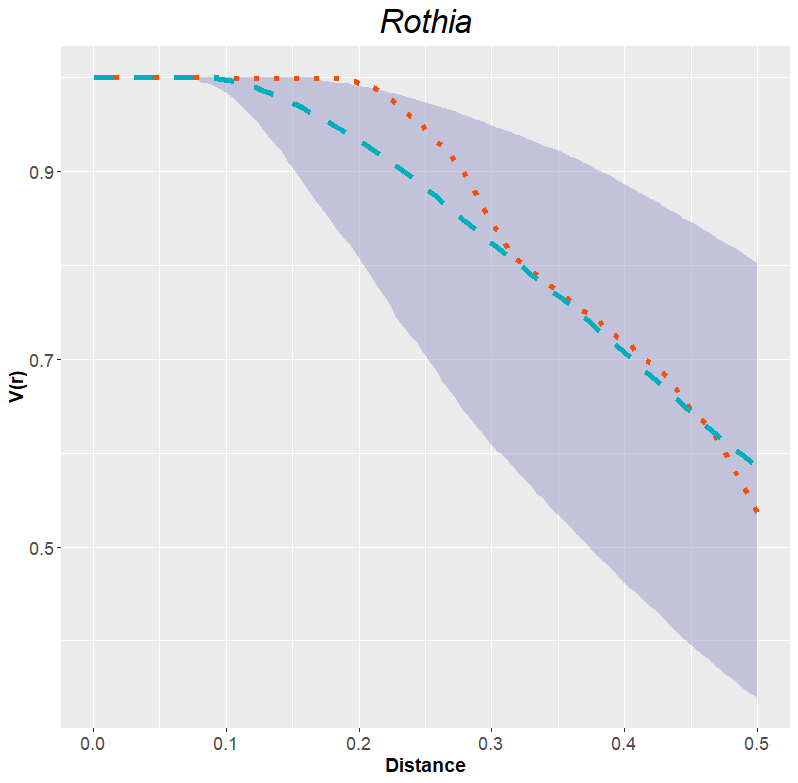}
         \caption{}
     \end{subfigure}
     \hfill
     \begin{subfigure}[b]{0.32\textwidth}
         \includegraphics[width=\textwidth]{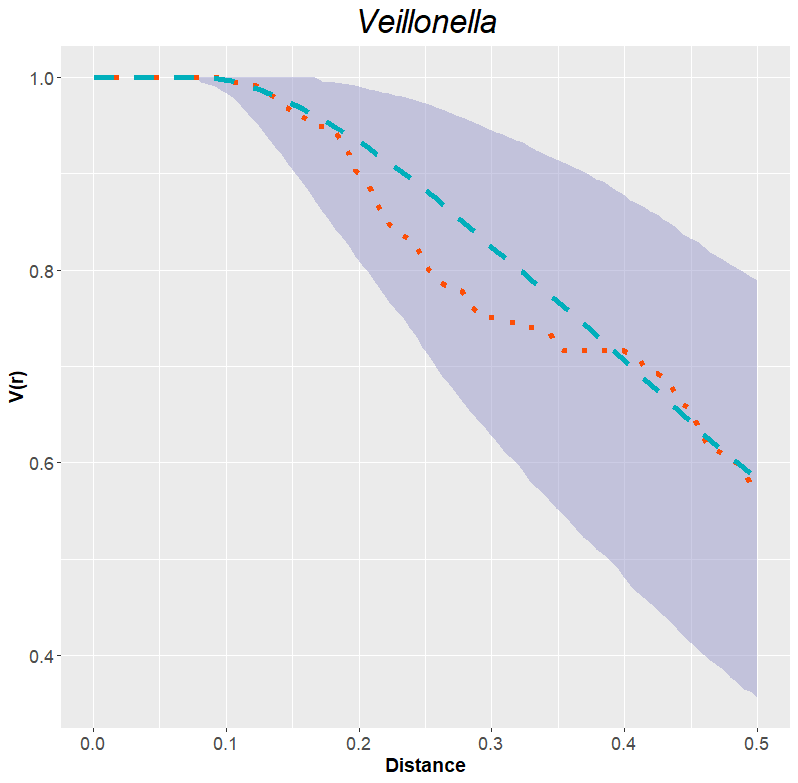}
         \caption{}
     \end{subfigure}
     \hfill
     \begin{subfigure}[b]{0.32\textwidth}
         \includegraphics[width=\textwidth]{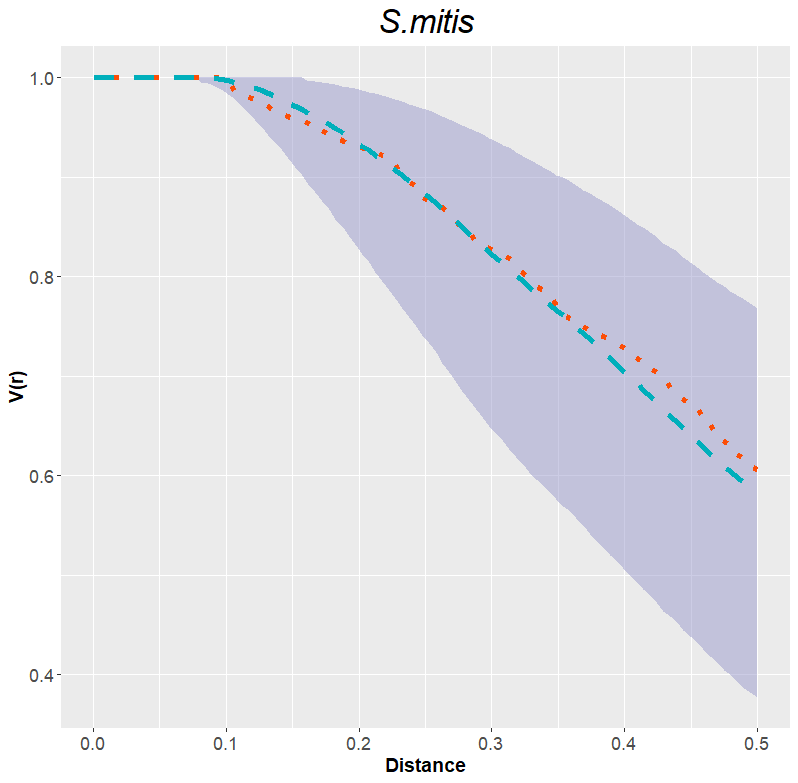}
         \caption{}
     \end{subfigure}
     \begin{subfigure}[b]{0.32\textwidth}
         \includegraphics[width=\textwidth]{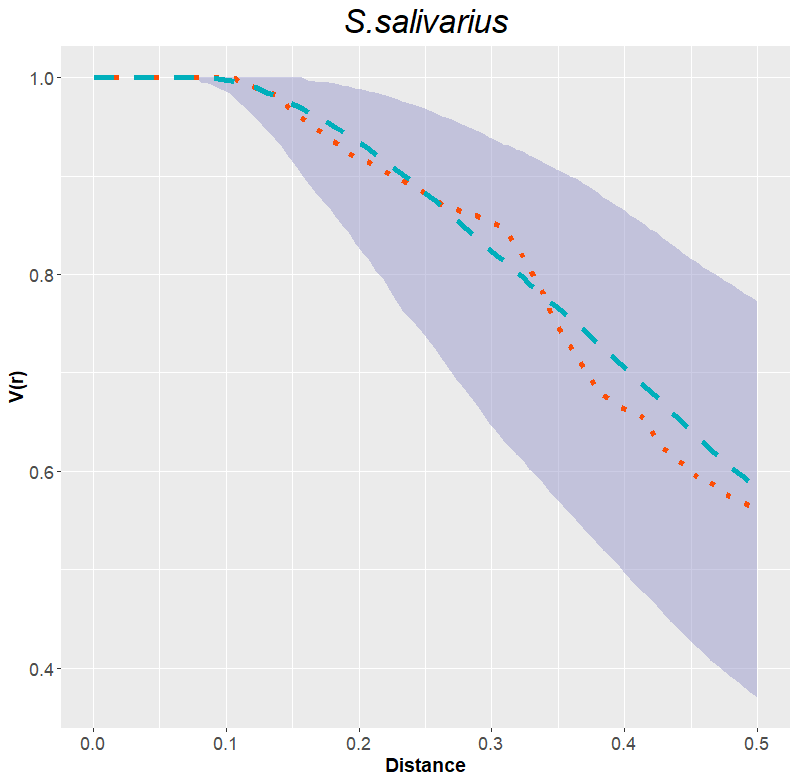}
         \caption{}
     \end{subfigure}
     \hfill
        \caption{Zero-probability function envelope plots for assessing posterior predictive distribution and model validation for each isolated taxa on the preserved tongue consortia from microbiome biofilm image B. The function estimates from the observations (red dotted curve), the point-wise means (blue dashed curve) and 95$\%$ interval (shaded area) obtained from 10,000 simulated posterior predictions were summarized for each isolated taxa.} 
\end{figure}

\begin{figure}
     \centering
     \begin{subfigure}[b]{0.32\textwidth}
         \includegraphics[width=\textwidth]{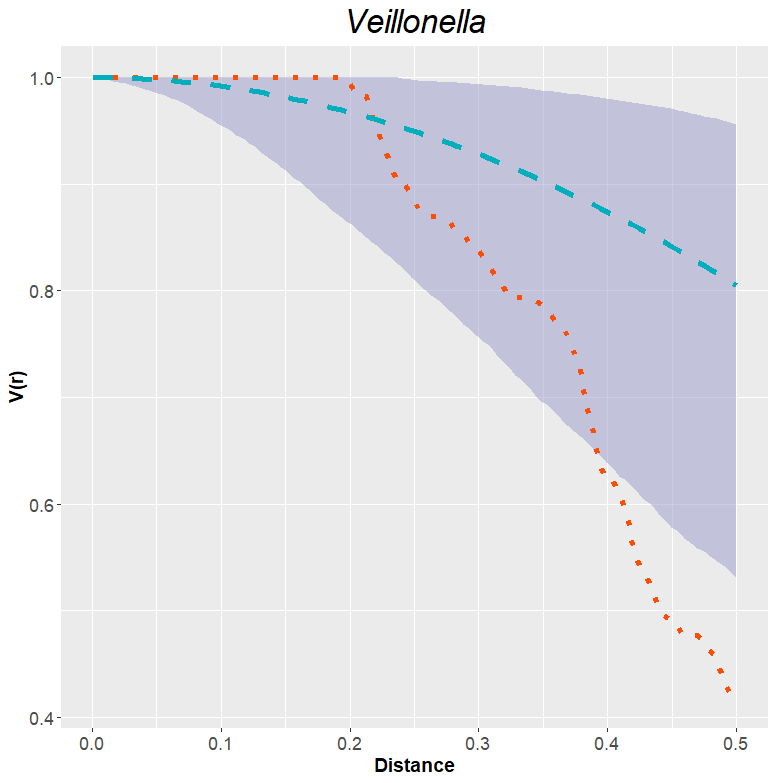}
         \caption{}
     \end{subfigure}
     \begin{subfigure}[b]{0.33\textwidth}
         \includegraphics[width=\textwidth]{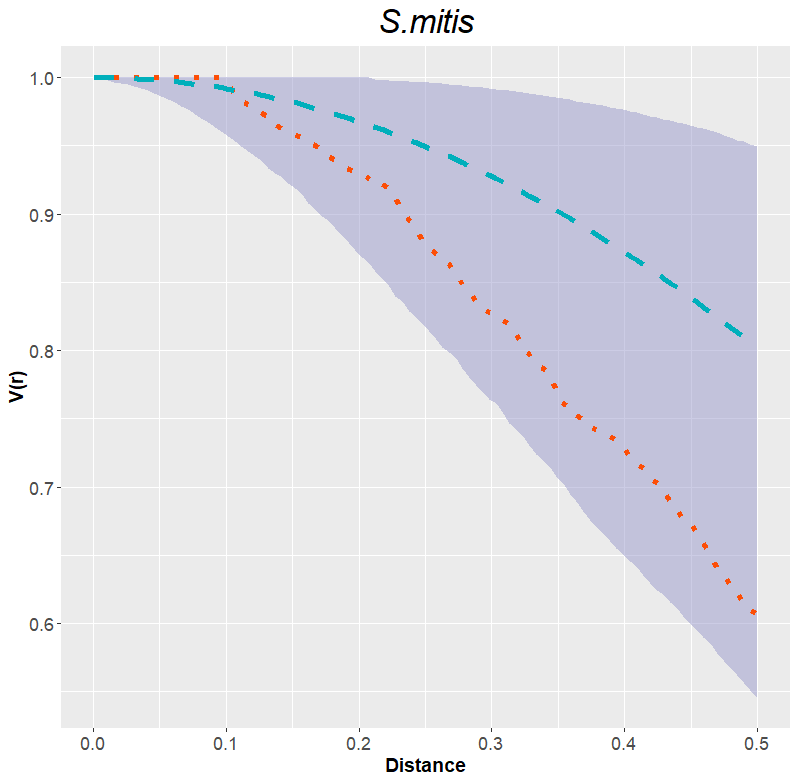}
         \caption{}
     \end{subfigure}
     \hfill
        \caption{Examples of zero-probability function envelope plot for the evaluation of spatial distribution regardless the preserved tongue consortia for selected taxon from each image (i.e., \textit{Veillonella} from image A and \textit{Streptococcus mitis} from image B). The zero-probability function envelope plots were obtained from 10,000 realizations, respectively. (a) $\&$ (b): The red dotted curve (the zero-probability function estimates from the observations) hardly lay within shaded area (95$\%$ simulated interval), indicating biased estimates obtained.}
\end{figure}

\section{\textbf{Concluding remarks}}

Performing Bayesian statistical inference for spatial point processes is challenging as it not only requires careful fine-tuning to achieve accurate posterior estimates but also needs to accommodate high-dimensional latent spatial processes, which can be computationally expensive. Our implementation of \textit{BayesFlow} explores and illustrates the exciting possibility that adopting modern deep learning neural inference can mitigate some of the challenges of performing inference for spatial point processes.

INNs are adopted for the approximation of the posterior distribution of the parameters of interest. We have provided simulation investigations and a real-world biofilm image application as an illustrative example, both of which demonstrate the method's utility and robustness in performing likelihood-free inference for point patterns in the same model family. \par
The only necessary information about the model is that it must lend itself fairly easily to simulation and have a relatively small parameter set. Meeting that requirement for LGCPs, the primary benefit of this method is that once training has been performed, the procedure can infer full posteriors based on any data set involving the same model family as the one used in the training phase. From the performance metrics we generated for 2-D LGCPs, the cost (in time) of training the neural network is recovered as long as there is more than one dataset for which Bayesian inference is desired. When making posterior inference for LGCPs with \textit{BayesFlow}, the procedure is ``finer-grid-friendly" in the sense that the computational gain increases with finer grid resolutions. In contrast, MCMC methods become increasingly inefficient with standard sampling algorithms as the grid resolution increases. In addition to the computational advantage, the point-wise posterior estimates from \textit{BayesFlow} are fairly reliable and the uncertainty of the estimates (credible intervals) is also provided.\par
We applied the proposed \textit{BayesFlow} framework to two distinct microbial biofilm images from \cite{wilbert2020spatial}. Specifically, we trained the neural network on simulated data sets, focusing on the well-defined perimeter of the consortia while excluding inner host area given each microbiome biofilm image, and leveraged the framework to perform the inference of spatial self-association for multiple taxa. We validated the derived estimates by constructing point-wise envelope plots of zero-probability function statistics. Furthermore, we also conducted the same validation procedure for the estimates from \textit{BayesFlow} when the regions outside microbial consortia as well as inner host area were not excluded when simulating data sets for the training phase. The findings agreed with previous studies (e.g., \citealp{schillinger2012co}); i.e.,  bias was introduced without distinguishing biomass-containing regions from the background.

 \par
The novel microbiome application integrating a deep learning framework for spatial distribution quantification also provides potential opportunities for researchers interested in characterizing spatial dependence structures among multiple taxa. One of the most widely used tools for examining spatial relationships in microbial ecology—the linear dipole algorithm implemented in the software $\texttt{daime}$ \citep{daims2006daime}—has notable limitations. For example, it lacks uncertainty quantification for single-image analysis, instead requiring multiple images to estimate empirical standard errors. As a more reliable alternative, \textit{BayesFlow} offers a model-driven framework that allows for rapid and stable statistical inference with straightforward interpretation. \par

We applied the proposed method to images of microbial communities as an illustrative example, highlighting its potential across a wide range of disciplines. We anticipate that method's advantages will be especially evident when applied to datasets with universally constrained boundaries, such as brain neuroimaging data, and to biofilm images involving a greater number of identified taxa, where the amortized inference feature is expected to offer even more exceptional computational gains. With further development, this deep-learning method can serve as a powerful alternative for accelerated model-based spatial statistical inference, benefiting a broader range of research communities. \par

\section*{Acknowledgments and Funding}

Shuwan Wang, Jessica L. Mark Welch, Jacqueline R. Starr, and Kyu Ha Lee were supported by the National Institute of Dental and Craniofacial Research (R21DE026872) and the National Institute of General Medical Sciences (R01GM126257).

\bibliographystyle{apalike}
\bibliography{bibtex_entries}

\end{document}